\documentclass[twocolumn, preprintnumbers,prb,aps,amssymb,showpacs,superscriptaddress]{revtex4}
\pdfoutput=1
\usepackage{graphicx}
\usepackage{dcolumn}
\usepackage{bm} \usepackage{color}

\begin{document}
%\preprint{4p0mt}

\newcommand{\ie}{{\it i.e.}}
\newcommand{\eg}{{\it e.g.}}
\newcommand{\etal}{{\it et al.}}

%%%%%%%%%%%%%%%%%%%%%%%%%%%% TITLE

\title{Nodes in the gap structure of the iron-arsenide superconductor
Ba(Fe$_{1-x}$Co$_x$)$_2$As$_2$ from $c$-axis heat transport measurements}

%%%%%%%%%%%%%%%%%%%%%%%%%%%% AUTHORS

\author{J.-Ph.~Reid}
\affiliation{D\'epartement de physique
\& RQMP, Universit\'e de Sherbrooke, Sherbrooke, Qu\'ebec, Canada}

\author{M.~A.~Tanatar}
\affiliation{Ames Laboratory, Ames, Iowa 50011, USA}

\author{X.~G.~Luo}
\affiliation{D\'epartement de physique
\& RQMP, Universit\'e de Sherbrooke, Sherbrooke, Qu\'ebec, Canada}

\author{H.~Shakeripour}
\affiliation{D\'epartement de physique
\& RQMP, Universit\'e de Sherbrooke, Sherbrooke, Qu\'ebec, Canada}

\author{N.~Doiron-Leyraud} 
\affiliation{D\'epartement de physique
\& RQMP, Universit\'e de Sherbrooke, Sherbrooke, Qu\'ebec, Canada}

\author{N.~Ni}
\affiliation{Ames Laboratory, Ames, Iowa 50011, USA}
\affiliation{Department of Physics and Astronomy, Iowa State
University, Ames, Iowa 50011, USA }

\author{S.~L.~Bud'ko}
\affiliation{Ames Laboratory, Ames, Iowa 50011, USA}
\affiliation{Department of Physics and Astronomy, Iowa State
University, Ames, Iowa 50011, USA }

\author{P.~C.~Canfield}
\affiliation{Ames Laboratory, Ames, Iowa 50011, USA}
\affiliation{Department of Physics and Astronomy, Iowa State
University, Ames, Iowa 50011, USA }

\author{R.~Prozorov}
 \affiliation{Ames Laboratory, Ames,
Iowa 50011, USA} \affiliation{Department of Physics and Astronomy,
Iowa State University, Ames, Iowa 50011, USA }

\author{Louis Taillefer}
\altaffiliation{E-mail: louis.taillefer@physique.usherbrooke.ca }
\affiliation{D\'epartement de physique \& RQMP, Universit\'e de
Sherbrooke, Sherbrooke, Qu\'ebec, Canada} \affiliation{Canadian Institute for
Advanced Research, Toronto, Ontario, Canada}

\date{\today}

%%%%%%%%%%%%%%%%%%%%%%%%%%%% ABSTRACT

\begin{abstract}

The thermal conductivity $\kappa$ of the iron-arsenide superconductor Ba(Fe$_{1-x}$Co$_x$)$_2$As$_2$ was measured down to 50~mK
for a heat current parallel ($\kappa_c$) 
and perpendicular ($\kappa_a$) to the tetragonal $c$ axis, for seven Co concentrations from underdoped to overdoped regions of the phase diagram 
($0.038 \leq x \leq 0.127$). 
A residual linear term $\kappa _{c0}/T$ is observed in the $T \to 0$ limit when the current is along the $c$ axis,
revealing the presence of nodes in the gap.
Because the nodes appear as $x$ moves away from the concentration of maximal $T_c$, they must be accidental, not imposed by symmetry, 
and are therefore compatible with an $s_{\pm}$ state, for example.
The fact that the in-plane residual linear term $\kappa _{a0}/T$ is negligible at all $x$ implies that 
the nodes are located in regions of the Fermi surface that contribute strongly to $c$-axis conduction and very little to in-plane conduction.
Application of a moderate magnetic field ({\it e.g.} $H_{\rm c2}/4$) excites quasiparticles that conduct heat along the $a$~axis just
as well as the nodal quasiparticles conduct along the $c$~axis.
This shows that the gap must be very small (but non-zero) in regions of the Fermi surface which contribute significantly to in-plane conduction.
These findings can be understood in terms of a strong ${\bf k}$ dependence of the gap $\Delta({\bf k})$ 
which produces nodes on a Fermi surface sheet with pronounced $c$-axis dispersion and deep minima on the remaining, quasi-two-dimensional sheets.

\end{abstract}

%74.25.Fy Transport properties (electric and thermal conductivity, thermoelectric effects, etc.)
%74.20.Rp Pairing symmetries (other than s-wave)
%74.70.Dd Ternary, quaternary, and multinary compounds (including Chevrel phases, borocarbides, etc.)

\maketitle

%%%%%%%%%%%%%%%%%%%%%%%%%%%%  INTRODUCTION  %%%%%%%%%%%%%%%%%%%%%%%%%%%%%%%%%%%%%%%%%%%%%%%%%%%%%%%%%%%%%%%%%%%%%

\section{Introduction}

%The concept of a superconducting gap, $\Delta$, is central to the field of superconductivity. The gap opening reduces the total electronic energy of a metal and makes the %uperconducting state stable. This is why the establishement of the $d$-wave symmetry of the gap function in the cuprates was such a crucial advancement %\cite{cupratesdwave}, showing that superconductivity with high $T_c$ is principally different from that found in usual metals. Many theories view this difference as a %reflection of a non-phonon pairing mechanism \cite{Monthoux,mathur}. 

\begin{figure} [t]
\centering
\includegraphics[width=8.5cm]{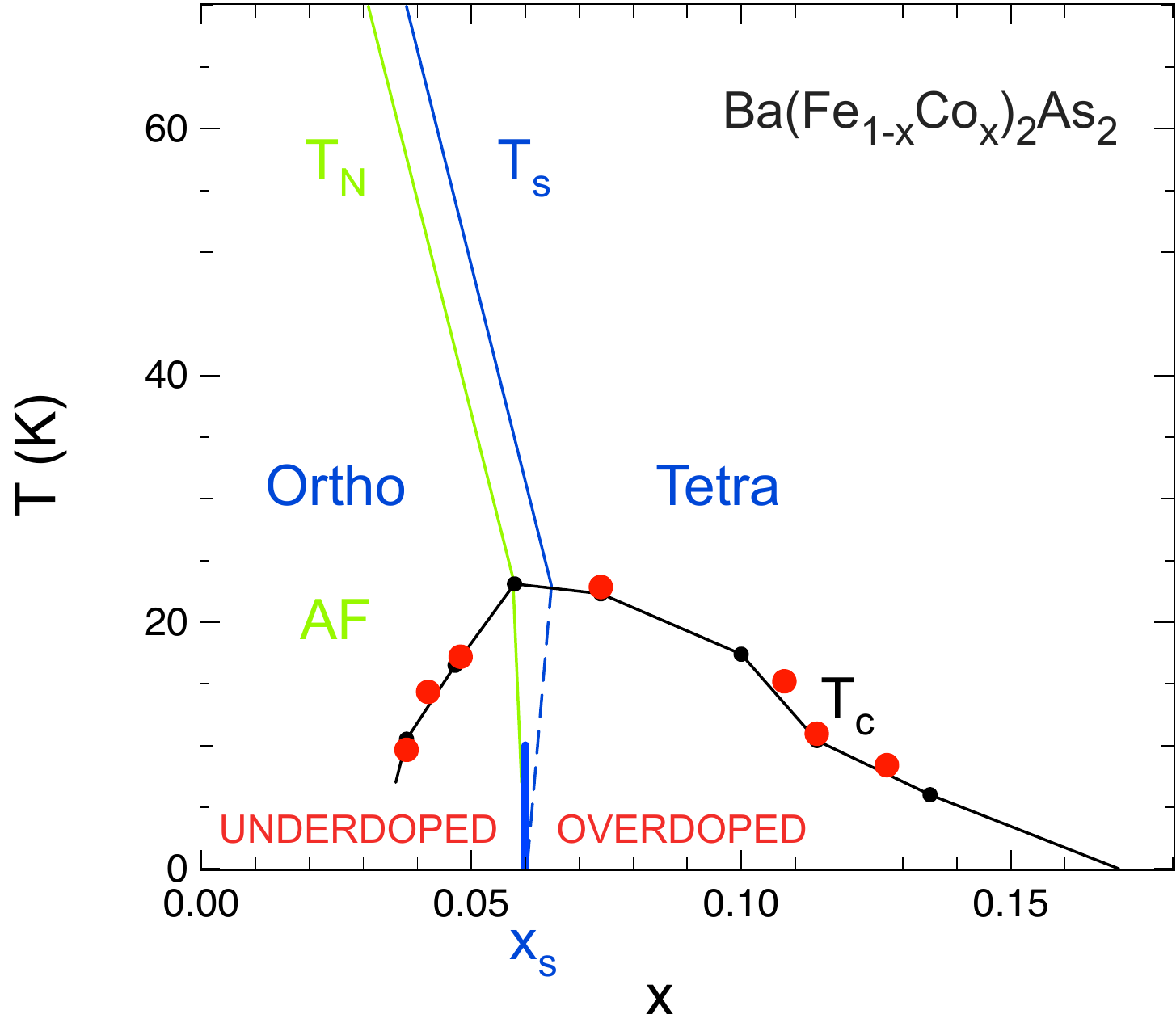}
\caption{
Phase diagram of Ba(Fe$_{1-x}$Co$_x$)$_2$As$_2$ as a function of Co concentration $x$, showing the orthorhombic phase below $T_s$ (blue), 
the antiferromagnetic (AF) phase below $T_N$ (green), 
and the superconducting phase below $T_c$ (small black dots),  
as determined from resistivity, magnetization and heat capacity data,\cite{Ni2008} and from x-ray~\cite{Nandi2010} and 
neutron~\cite{Fernandes2010} data.
Red circles mark the $T_c$ value of seven samples measured in the present study (the $c$-axis samples labelled A in Table~I), indicating the range of concentrations covered.
The vertical dashed line at $x = 0.06$ marks the approximate location of the critical concentration $x_s$ where at $T=0$ the system goes from orthorhombic (Ortho) in the underdoped region (to the left) to tetragonal (Tetra) in the overdoped region (to the right).\cite{Nandi2010}
At $T=0$, the AF phase also ends close to $x = 0.06$.\cite{Fernandes2010}
}
\label{phase-diagram}
\end{figure}

The discovery of superconductivity in iron arsenides,\cite{Kamihara2008} with transition temperatures exceeding 50~K,\cite{Zhi-An2008} breaks the monopoly of cuprates as the only family of high-temperature superconductors, and revives the question of the pairing mechanism.
Because the mechanism is intimately related to the symmetry of the order parameter, which is in turn related to the ${\bf k}$ dependence of the gap function 
$\Delta({\bf k})$, it
is important to determine the gap structure in the iron-based superconductors, just as it was crucial to establish the $d$-wave symmetry of the gap in cuprate superconductors.
The gap structure of iron-based superconductors has been the subject of numerous studies (for recent reviews, see Refs.~\onlinecite{Ishida2009,Mazin2010}). 
Here, we focus on the material BaFe$_2$As$_2$, 
in which superconductivity can be induced either by applying pressure~\cite{Alireza2009} or by various chemical substitutions, such as
K for Ba (K-Ba122)~\cite{Rotter2008} or Co for Fe (Co-Ba122).\cite{Sefat2008} 
In the case of Co-Ba122,
single crystals have been grown with compositions that cover the entire superconducting phase 
(see Fig.~\ref{phase-diagram}).\cite{Ni2008, Nandi2010, Fernandes2010}  

Two sets of experiments on doped BaFe$_2$As$_2$ appear to give contradictory information.
On the one hand, angle-resolved photoemission spectroscopy (ARPES) detects a nodeless, isotropic superconducting gap on all sheets of the Fermi surface in K-Ba122 \cite{Nakayama2009} and in optimally-doped Co-Ba122,\cite{Terashima2009} and tunneling studies in K-Ba122 detect two full superconducting gaps.\cite{Samuely2009} 
The magnitude of the gaps in ARPES is largest on Fermi surfaces where a density-wave gap develops in the parent compounds.\cite{Xu2009} 
This is taken as evidence for an s$_{\pm}$ pairing state driven by antiferromagnetic correlations.\cite{Mazin2008,Kuroki2008,Vorontsov2008,Mazin2010} 
On the other hand, the penetration depth in Co-Ba122,\cite{Gordon2009a,Gordon2009} the spin-lattice relaxation rate in K-Ba122,\cite{Fukazawa2009} and the in-plane thermal conductivity in K-Ba122 \cite{Luo2009} and Co-Ba122,\cite{Tanatar2010,Dong2010} for example, are inconsistent with a gap that is large everywhere on the Fermi surface. 
Note, however, that the evidence for deep minima in the gap is particularly clear in the overdoped regime,\cite{Fukazawa2009,Tanatar2010,Martin2010} a regime which has not so far been probed by either ARPES or tunneling.

Another possible explanation for the apparent discrepancy between the two sets of experimental results is a different sensitivity to 
the $c$-axis component ($k_z$) of the quasiparticle ${\bf k}$ vector, taking into account the three-dimensional (3D) character of the Fermi surface. \cite{Sefat2008,Malaeb2009,Utfeld2010,Kemper2009,Analytis2009,Tanatar2009,Tanatar2009a}
Nodes along the $c$~axis were suggested theoretically to explain the discrepancy between ARPES, penetration depth and NMR studies.\cite{Laad2009,Graser2010} 
A variation of the gap magnitude as a function of $k_z$ was suggested 
in experimental studies of the neutron resonances in optimally-doped Ni-Ba122.\cite{Chi2009} 
It was also invoked to explain the temperature dependence of the penetration depth in Co-Ba122~\cite{Gordon2009} and its $a-c$ anisotropy in
Ni-Ba122.\cite{Martin2010} 
Clearly, it has become important to resolve the 3D structure of the superconducting gap function in doped BaFe$_2$As$_2$.

Heat transport measured at very low temperatures is one of the few directional bulk probes of the gap structure.
The existence of a finite residual linear term $\kappa_0/T$ in the thermal conductivity $\kappa(T)$ as $T \to 0$ is unambiguous evidence for the presence of nodes in the gap,\cite{Hirschfeld1988,Graf1996,Durst2000,Taillefer1997,Suzuki2002} and thus
by measuring $\kappa(T)$ as a function of direction in the crystal, one can locate the position of nodes on the Fermi surface.\cite{Hirschfeld1988,Graf1996,Shakeripour2009a,Shakeripour2007}
Here, we report measurements of heat transport in Ba(Fe$_{1-x}$Co$_x$)$_2$As$_2$ for a current direction both parallel and perpendicular to the $c$ axis of the tetragonal (or orthorhombic) crystal structure.  
Our main finding is a sizable residual linear term $\kappa_0/T$ for a current along the $c$ axis, and a negligible one for a current perpendicular to it. 
This implies the presence of nodes in the gap in regions of the Fermi surface that dominate the $c$-axis conduction and contribute little to in-plane conduction. 
Our study shows that the gap structure of Co-Ba122 depends on the 3D character of the Fermi surface in a way that varies strongly with $x$.

%------------------ FIGURE 1 --------------------------------------------------------------------------------------------------------

%------------------------------------------------------------------------------------------------------------------------------------

%------------------ FIGURE 2 --------------------------------------------------------------------------------------------------------

\begin{figure} [t]
\centering
\includegraphics[width=8.5cm]{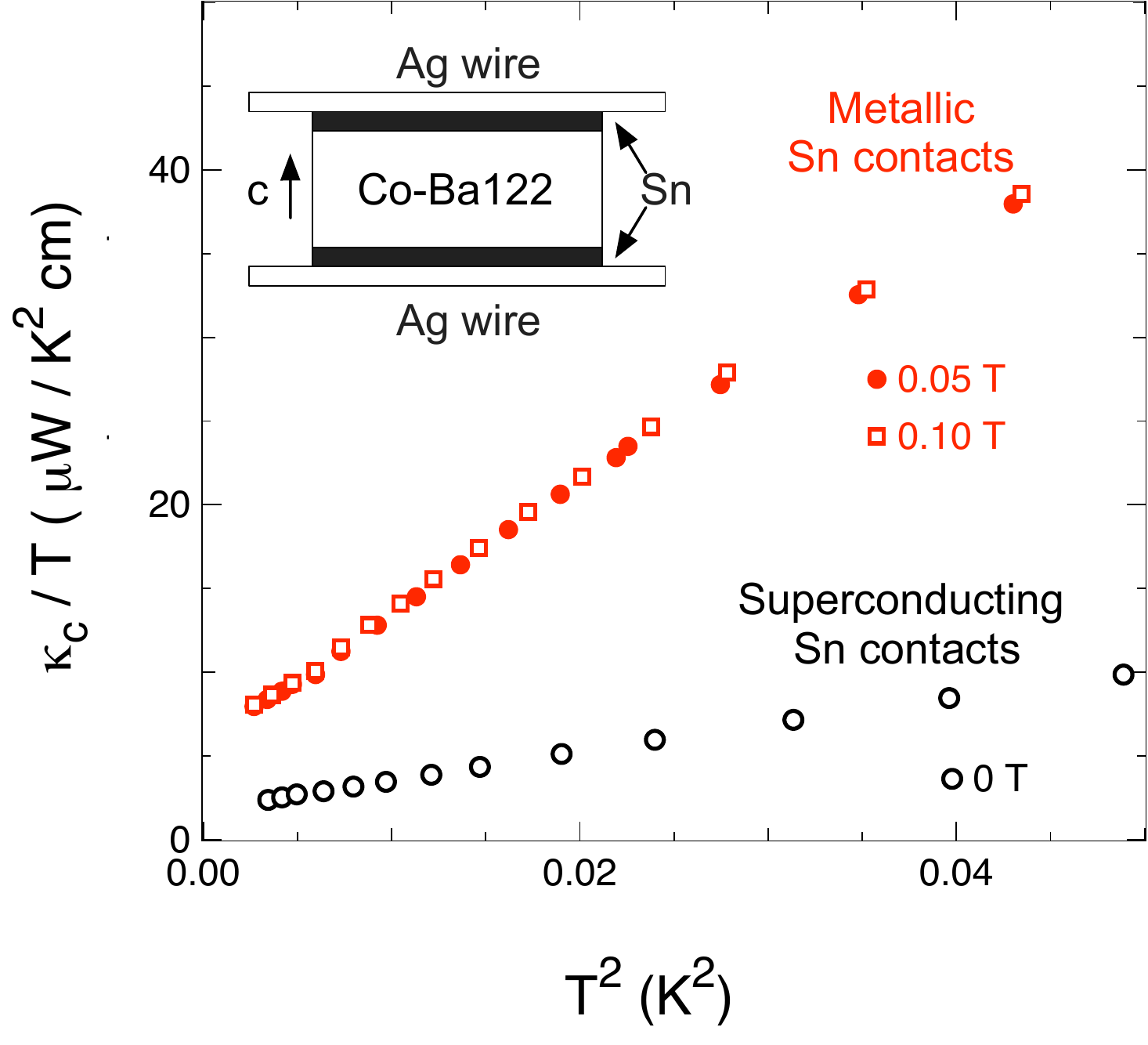}
\caption{
Temperature dependence of the thermal conductivity $\kappa$, plotted as $\kappa/T$ vs $T^2$, measured along the $c$~axis in a Co-Ba122 sample with $x=0.127$,
using the two-probe technique, for three values of the applied field: $H = 0.0$, 0.05 and 0.10~T.
At $H=0$, the total thermal resistance of sample plus contacts is dominated by the very high thermal resistance of the two superconducting tin contacts. 
At $H=50$~mT, tin is no longer superconducting and the thermal resistance of the two contacts has become negligible compared to the sample resistance. 
In this case, extrapolation of $\kappa/T$ to $T$=0 gives us almost exactly the residual linear term in the sample's thermal conductivity. 
Increasing $H$ slightly beyond 50~mT, for example to 100~mT, leads to no further change in the data.
This shows that measurements in $H = 50$~mT reveal the intrinsic zero-field behavior of the sample.
The inset shows the arrangement of the tin (Sn) contacts and the silver (Ag) wires on the sample. 
}
\label{thermal_contact_resistance}
\end{figure}

%--------------------------------------------------------------------------------------------------------------------------------------

%------------------ TABLE I --------------------------------------------------------------------------------------------------------

\begin{table}[t]
\caption{Properties of the twelve $c$-axis samples of Co-Ba122 used in this study. 
$x$ is the Co concentration measured by wavelength dispersive microprobe analysis.
The superconducting transition temperature $T_c$ is the temperature at which the resistivity goes to zero.
The values of the upper critical field $H_{c2}$ needed to suppress superconductivity in Co-Ba122 at $T \to 0$ are taken from Refs.~\onlinecite{Ni2008}
and~\onlinecite{Kano2009}.
($H_{c2}(T)$ is defined as the end, or `offset', of the superconducting drop in $\rho(T)$ vs $H$).
The residual resistivity $\rho_{c0}$ is obtained by a smooth extrapolation of the $\rho_c(T)$ data to $T=0$, as shown in Fig.~\ref{rhocT}.
The normal-state residual linear term in the thermal conductivity, $\kappa_{\rm cN}/T$, is obtained from the Wiedemann-Franz law applied to $\rho_{c0}$ (see text).
The zero-field residual linear term, $\kappa_{c0}/T$, is obtained by extrapolating to $T=0$ the zero-field thermal conductivity $\kappa_c$ with a linear fit to $\kappa_c/T$ vs $T^2$, as shown in Fig.~\ref{kappacoverTvsT2}. 
$\kappa_{c0}/T$ is also expressed as a fraction of the normal-state $\kappa_{\rm cN}/T$, denoted $\kappa_0/\kappa_{\rm N} \equiv (\kappa_{\rm 0}/T)/(\kappa_{\rm N}/T)$.
%
%For five Co concentrations, two different crystals with nominally the same $x$ value, labelled A and B, were measured.
%Only the raw data for the A samples are displayed in Fig.~\ref{kappaTvsT2andkappavsH-underdoped} and Fig.~\ref{kappaTvsT2andkappavsH-overdoped};
%for all 12 samples, $\kappa_{c0}/T$ is plotted vs $H$ in Fig.~\ref{reproducibility},
%and $\kappa_0/\kappa_{\rm N}$ is plotted vs $x$ in Fig.~\ref{summary}.
%
%The uncertainty on geometric factors confers a $\pm 30 \%$ error bar on $\rho_{c0}$, $\kappa_{\rm cN}/T$, and $\kappa_{c0}/T$.
%This uncertainty cancels out in the ratio $\kappa_0/\kappa_{\rm N}$ because the same contacts are used for thermal and electrical measurements.
%This leaves a $\pm 5 \%$ error bar on $\kappa_{\rm cN}/T$ from the extrapolation of $\rho_{c0}$, and a $\pm 0.5~\mu$W/K$^2$~cm error bar on
%$\kappa_{c0}/T$.
}
\centering
	\begin{tabular}{|c | c | c | c| c|  c| c| c|}
		\hline
$x$ & Sample &	$T_{c}$ & $H_{c2}$ &  $\rho_{c0}$ &  $\kappa_{\rm cN}/T $	 & $\kappa_{c0}/T$ & $\kappa_0/\kappa_{\rm N}$	\\
 & & (K) & (T) & $(\mu\Omega$~cm) & \multicolumn{2}{c|}{($\mu$W/K$^2$~cm)} &	 \\
		\hline\hline
%			      T_c		Hc2		 rho	KappaN	Kc0		 K0/Kn
0.038	 &A	 & 9.7	&30		&1935	 &12.7	&6.1   &0.48			\\			%10536
\hline
0.042	 &A	 &14.4	&40		&1980	 &12.4	&2.3   &0.19			\\			%9263
0.042	 &B	 &13.7	&40		&2115	 &11.6	&2.9   &0.25			\\			%10542
\hline
0.048	 &A	 &17.2	&45		&2535	 & 9.7	&0.6   &0.06			\\			%9456
0.048	 &B	 &17.2	&45		&3045	 & 8.0	&0.8   &0.10			\\			%10548
\hline
0.074  &A  &22.9	&60 	&1030  &23.8	&0.9   &0.04			\\			%9247
0.074  &B  &24.1	&60 	&1140  &21.5	&0.2   &0.01			\\			%9202
\hline
0.108  &A  &15.2	&30		&1560	 &15.7	&2.3   &0.15			\\			%9208
0.108  &B  &14.6	&30		&1770	 &13.8	&1.6   &0.12			\\			%9209
\hline
0.114	 &A  &11.0 	&20 	&1415	 &17.3	&3.8   &0.22			\\			%9224
\hline
0.127  &A  & 8.4 	&15		&1500	 &16.3	&5.6   &0.34			\\			%9270
0.127  &B  & 9.3 	&15		&1130	 &21.7	&6.8   &0.31			\\		 	%9271
		\hline
	\end{tabular}
	
\label{table1}
\end{table}

%--------------------------------------------------------------------------------------------------------------------------------------

%%%%%%%%%%%%%%%%%%%%%%%%%%%%  EXPERIMENTAL  %%%%%%%%%%%%%%%%%%%%%%%%%%%%%%%%%%%%%%%%%%%%%%%%%%%%%%%%%%%%%%%%%%%%%

\section{Experimental}

\subsection{Samples}

Single crystals of Ba(Fe$_{1-x}$Co$_x$)$_2$As$_2$ were grown from FeAs:CoAs flux, as described elsewhere.\cite{Ni2008} 
The doping level in the crystals was determined by wavelength dispersive electron probe microanalysis, 
which gave a Co concentration, $x$, roughly 0.7 times the flux load composition (or nominal content). 
We studied seven compositions: 
underdoped, with $x = 0.038$, 0.042, and 0.048; 
overdoped, with $x=0.074$, 0.108, 0.114, and 0.127.
In this Article, `underdoped' and `overdoped' refer to concentrations respectively below and above the critical concentration 
$x_s \simeq 0.06$ at which the system at $T=0$ goes from orthorhombic (below) to tetragonal (above).\cite{Nandi2010}
The $T_c$ value for each composition is shown on the phase diagram in Fig.~\ref{phase-diagram}.
A total of twelve $c$-axis and nine $a$-axis samples were studied; their characteristics are listed in Tables~I and II, respectively. 
Three of the $a$-axis samples were the subject of a previous study (0.074-B, 0.108-A, and 0.114-B).\cite{Tanatar2010}

\subsection{Two-probe transport measurements}

Thermal conductivity was measured in a standard one-heater-two-thermometer technique.\cite{Sutherland2003}
The magnetic field $H$ was applied along the [001] or $c$-axis direction of the crystal structure, which is tetragonal for overdoped samples and orthorhombic for underdoped samples at low temperatures. 
Data were taken on warming after having cooled in a constant field applied above $T_c$ to ensure a homogeneous field distribution.

It is conventional to measure electrical and thermal resistance in a four-probe configuration to avoid the contribution of contact resistances. 
This is what was done for data taken with a current in the basal plane ($J~||~a$, in the notation appropriate for the tetragonal phase), 
as described elsewhere.\cite{Luo2009,Tanatar2010}
For a current along the $c$ axis ($J~||~c$), however, the four-probe technique is difficult because of the strong tendency of iron-arsenide crystals to exfoliation, 
which makes it difficult to cut samples thick enough in the $c$ direction to attach four contacts.\cite{Tanatar2009,Tanatar2009a}
Consequently, $c$-axis transport was measured using a two-probe technique, which is valid provided contact resistances are much smaller than the sample resistance.

Contacts to the $c$-axis samples were made using silver wires (of 50 $\mu$m diameter), soldered to the top and bottom surfaces of the sample with ultrapure tin (see inset of Fig.~\ref{thermal_contact_resistance}). The contact making and properties are described in detail in Ref.~\onlinecite{Tanatar2010a}. In brief, these contacts are characterized by a surface area resistivity in the 
n$\Omega$~cm$^2$ range, which, for a typical sample size, yields a contact resistance below 10 $\mu \Omega$. This is negligible compared to a typical sample resistance in the normal state, of the order of 10~m$\Omega$. 

Because tin is a superconductor, the thermal resistance of the contacts at very low temperature is large. 
We therefore have to apply a small magnetic field to suppress the superconductivity of tin and make it a normal metal, with the very low electrical and thermal resistance mentioned above. A field of 50 mT is sufficient to do this. 
In Fig.~\ref{thermal_contact_resistance}, we compare data obtained with $H = 0.00$, 0.05 and 0.10~T. The effect of switching off the contact resistance with the field is clear, and once tin has gone normal, the data is independent of a further small increase in $H$. We therefore regard the data taken at $H = 0.05$~T as representative of the zero-field state of the sample.

%------------------ TABLE II --------------------------------------------------------------------------------------------------------

\begin{table}[t]
\caption{Properties of the nine $a$-axis samples of Co-Ba122 used in this study. 
$x$, $T_c$, $H_{c2}$ and $\kappa_0/\kappa_{\rm N}$ are defined in Table~I.
The residual resistivity $\rho_{a0}$ is obtained by a smooth extrapolation of the $\rho_a(T)$ data to $T=0$, as shown in Fig.~\ref{rhoaT}.
The normal-state residual linear term in the thermal conductivity, $\kappa_{\rm aN}/T$, is obtained from the Wiedemann-Franz law applied to $\rho_{a0}$.
The zero-field residual linear term, $\kappa_{a0}/T$, is obtained by extrapolating to $T=0$ the zero-field thermal conductivity $\kappa_a$ with a power-law fit to $\kappa_a/T$ (see text), as shown in Fig.~\ref{kappaaoverTvsT}.
Note that the magnitude of $\kappa_{a0}/T$, whether positive or negative, is in all cases lower than the uncertainty in the extrapolation 
(see Fig.~\ref{kappa0overkappaNvsx}).
%
%For three Co concentrations, two different crystals with nominally the same $x$ value, labelled A and B, were measured.
%Only the raw data for the A samples are displayed in Fig.~\ref{kappaaTvsT2andkappavsH-underdoped} and Fig.~\ref{kappaaTvsT2andkappavsH-overdoped};
%for all 10 samples, $\kappa_{a0}/T$ is plotted vs $H$ in Fig.~\ref{kappaavsHTto0limit}.
%
%Data for samples 0.048-A, 0.074-B, 0.108-A and 0.114-B were reported in ref.~\cite{Tanatar2010} {\bf *** Is this correct?}.
%
%The uncertainty on geometric factors confers a $\pm 20 \%$ error bar on $\rho_{a0}$, $\kappa_{\rm aN}/T$, and $\kappa_{a0}/T$.
%This uncertainty cancels out in the ratio $\kappa_0/\kappa_{\rm N}$ because the same contacts are used for thermal and electrical measurements.
%This leaves a $\pm 5 \%$ error bar on $\kappa_{\rm aN}/T$ from the extrapolation of $\rho_{a0}$, and a $\pm 5~\mu$W/K$^2$~cm error bar on
%$\kappa_{a0}/T$.
}
\centering
	\begin{tabular}{|c | c | c | c| c|  c| c| c|}
		\hline
$x$ & Sample &	$T_{c}$ & $H_{c2}$ &  $\rho_{a0}$ &  $\kappa_{\rm aN}/T $	 & $\kappa_{a0}/T$ & $\kappa_0/\kappa_{\rm N}$	\\
 & & (K) & (T) & $(\mu\Omega$~cm) & \multicolumn{2}{|c|}{($\mu$W/K$^2$~cm)}&	 \\
		\hline\hline
%			      T_c		Hc2		 rho	KappaN	Ka0		 K0/Kn
0.042	 &A	 &13.0	&40		&200 	 &123		& 1   & 0.01	  \\			%10543
0.042	 &B	 &14.2	&40		&235	 &104		& 0   & 0    	  \\			%10544
\hline
0.048	 &A	 &16.7	&45		&150	 &163		& 2   & 0.01	  \\			%74     
\hline
0.074  &A  &22.2	&60 	& 62   &395		&-1   & 0		    \\			%107
0.074  &B  &22.2	&60 	& 82   &299		& 3   & 0.01	  \\			%108     *** in PRL 
\hline
0.108  &A  &14.8	&30		& 59	 &415		&-1   & 0       \\			%127     *** in PRL 
\hline
0.114	 &A  &10.8  &20 	& 59	 &415		&-9   &-0.02    \\			%98
0.114	 &B  &10.2  &20 	& 56	 &438		&-13  &-0.03    \\			%100     *** in PRL 
\hline
0.127  &A  & 8.2 	&15		& 48	 &510		&17   & 0.03    \\			%9268

	\hline
	\end{tabular}	
\label{table2}
\end{table}

%--------------------------------------------------------------------------------------------------------------------------------------

\subsection{Electrical resistivity}

\begin{figure} [t]
\centering
\includegraphics[width=8.5cm]{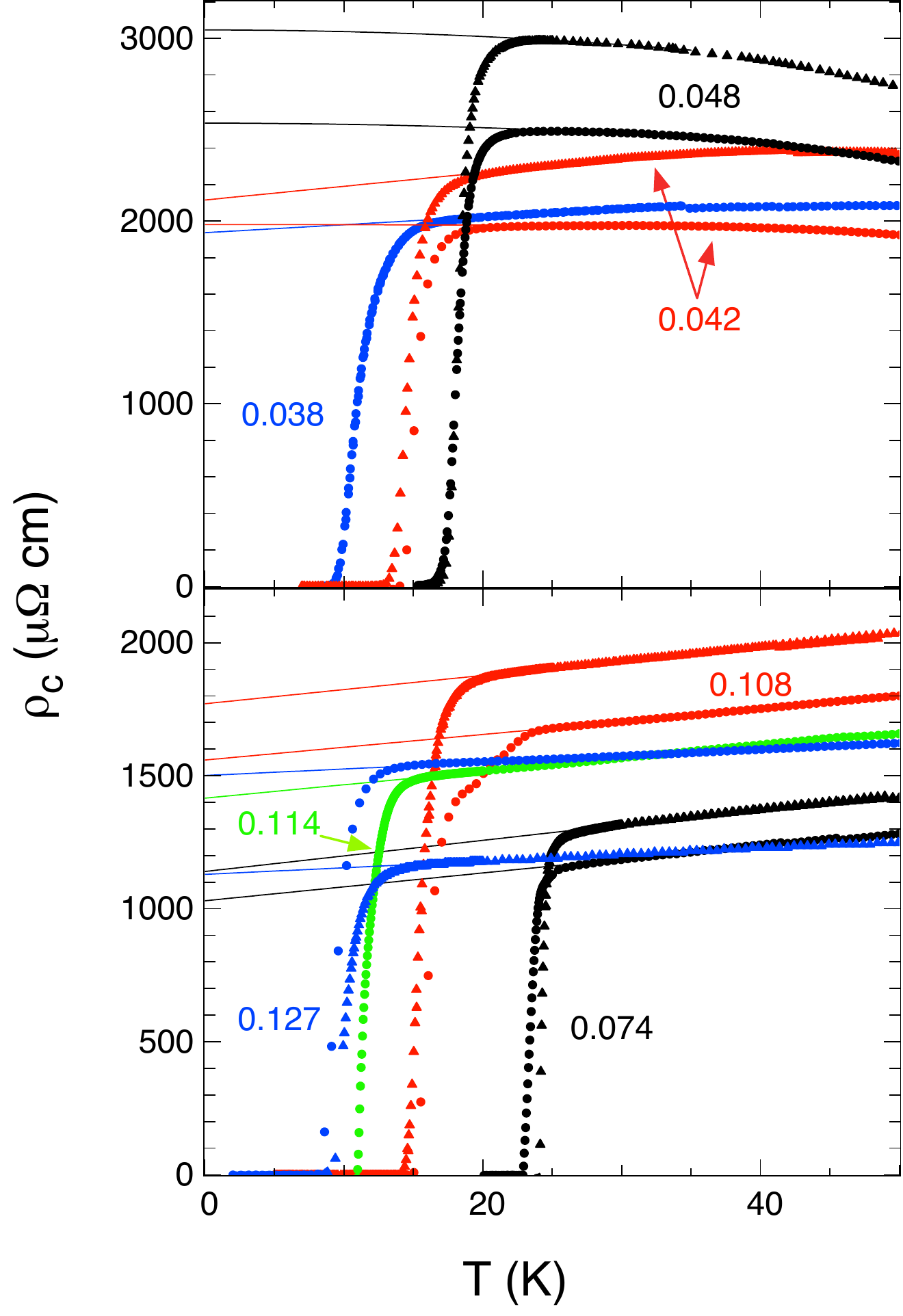}
\caption{
Temperature dependence of the $c$-axis resistivity $\rho_c(T)$ for the twelve $c$-axis crystals of Ba(Fe$_{1-x}$Co$_x$)$_2$As$_2$ studied here, 
with Co concentrations $x$ as indicated. 
Top and bottom panels show underdoped and overdoped samples, respectively. 
Circles and triangles of the same color correspond to two different crystals at the same doping, respectively labelled A and B (see Table~I).
%A difference in the absolute magnitude can come from the uncertainty in the geometric factor.
%
The lines show how the data is extrapolated to $T = 0$, to determine the value of the residual resistivity $\rho_{c0}$, 
given in Table~I. 
}
\label{rhocT}
\end{figure}

\begin{figure} [t]
\centering
\includegraphics[width=8.5cm]{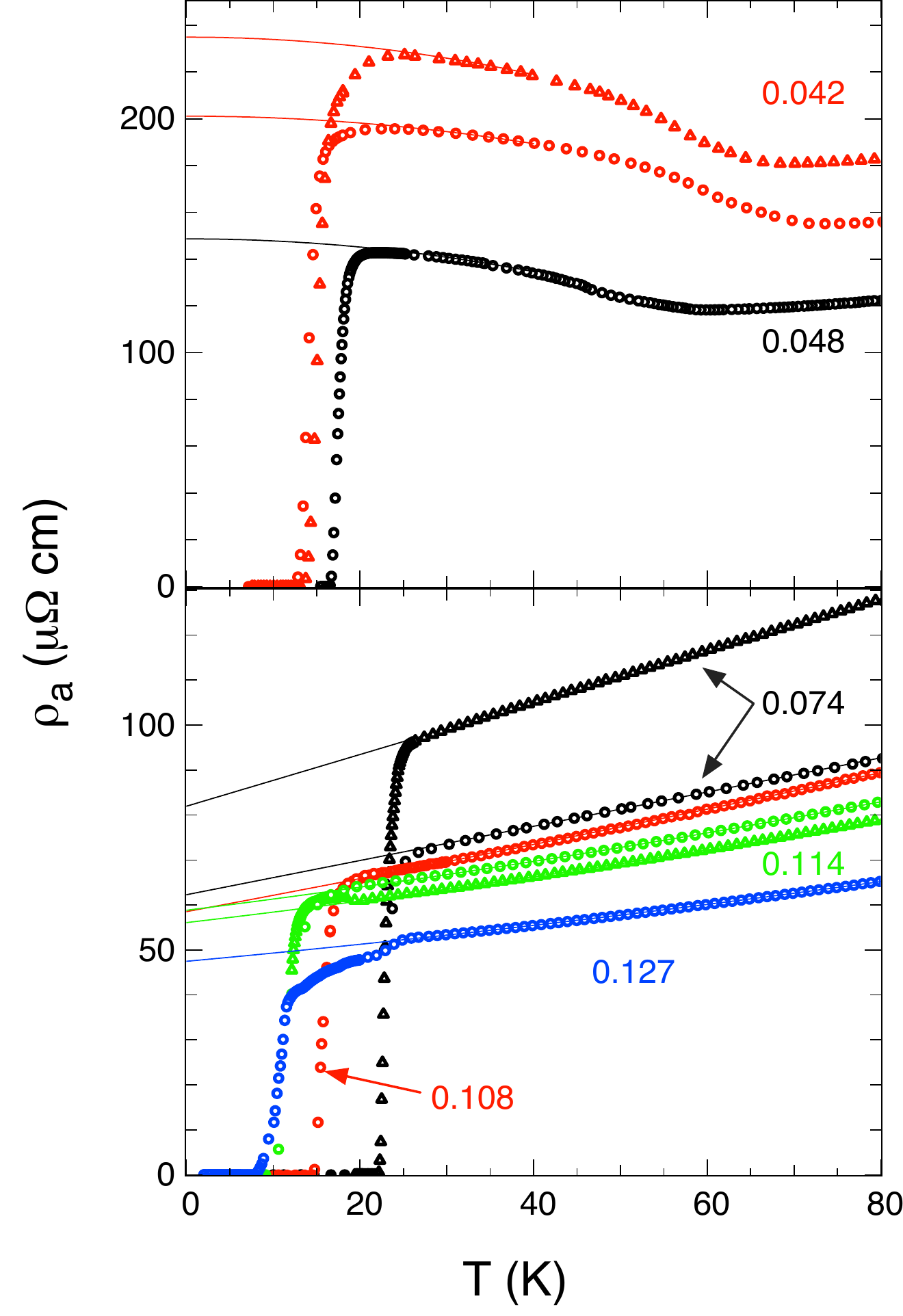}
\caption{
Temperature dependence of the $a$-axis (in-plane) resistivity $\rho_a(T)$ for the nine $a$-axis crystals of Ba(Fe$_{1-x}$Co$_x$)$_2$As$_2$ studied here, 
with Co concentrations $x$ as indicated. 
Top and bottom panels show underdoped and overdoped samples, respectively. 
Circles and triangles of the same color correspond to two different crystals at the same doping, respectively labelled A and B (see Table~II).
%A difference in the absolute magnitude can come from the uncertainty in the geometric factor.
%
The lines show how the data is extrapolated to $T = 0$, to determine the value of the residual resistivity $\rho_{a0}$, 
given in Table~II.}
\label{rhoaT}
\end{figure}

In Fig.~\ref{rhocT}, we show the temperature dependence of the electrical resistivity $\rho_c(T)$ of our $c$-axis samples, measured in a two-probe configuration.
%(Note the negligible value of the contact resistance measured between the $T_c$ of the sample (> 10 K) and the $T_c$ of tin (< 4 K).) 
In all samples, the resistivity follows qualitatively the temperature dependence reported previously.\cite{Tanatar2009} 
Data for the $a$-axis samples are shown in Fig.~\ref{rhoaT}.
A smooth extrapolation of $\rho(T)$ to $T = 0$ yields the residual resistivity $\rho_0$ listed in Table~I for $c$-axis samples and in Table~II for $a$-axis samples. 
The uncertainty associated with the extrapolation of $\rho(T)$ to $T = 0$ is approximately $\pm~5~\%$.
Due to the uncertainty in measuring the geometric factor, the absolute value of the resistivity has an error bar of approximately $\pm$~20~\% for $a$-axis samples and a factor of 2 uncertainty for $c$-axis samples. 
\cite{Tanatar2009,Tanatar2009a, Laad2009}
%
%For overdoped samples ($x > 0.065$), $\rho_{c0} = 1400 \pm 400~\mu\Omega$~cm and $rho_{a0} = 65 \pm 15~\mu\Omega$~cm, 
%while for underdoped samples ($x < 0.065$), $\rho_{c0} = 2500 \pm 500~\mu\Omega$~cm and $\rho_{a0} = 200 \pm 40~\mu\Omega$~cm.
The higher $\rho_0$ values in the underdoped regime are due to a reconstruction of the Fermi surface in the antiferromagnetic phase.
The residual resistivity $\rho _0$ is used to determine the normal-state thermal conductivity $\kappa_{\rm N}/T$ in the $T = 0$ limit via the Wiedemann-Franz law, $\kappa_{\rm N}/T = L_0 /\rho _0$, where $L_0 = 2.45~\times$~10$^{-8}$ W~$\Omega$~/~K$^2$.
Because the same contacts are used for electrical and thermal measurements, the relative geometric-factor uncertainty between the measured $\kappa$ and this electrically-determined $\kappa_{\rm N}$ is minimal.

%and a possible mismatch between heat and charge current path. In case when the sample is crack free, this difference comes from slight difference in geometric factors for %heat and charge flow in the contact areas and leads to approximately 10\% uncertainty in the Wiedemann-Franz law \cite{Science}.
%

%------------------ FIGURE 3 --------------------------------------------------------------------------------------------------------

%--------------------------------------------------------------------------------------------------------------------------------------

%------------------ FIGURE 4 --------------------------------------------------------------------------------------------------------

%--------------------------------------------------------------------------------------------------------------------------------------

%%%%%%%%%%%%%%%%%%%%%%%%%%%%  RESULTS  %%%%%%%%%%%%%%%%%%%%%%%%%%%%%%%%%%%%%%%%%%%%%%%%%%%%%%%%%%%%%%%%%%%%%

\section{Results}

\subsection{Heat transport in the $c$ direction} 

The thermal conductivity of solids is the sum of electronic and phononic contributions: $\kappa = \kappa_e + \kappa_p$. 
In the $T = 0$ limit, the electronic conductivity is linear in temperature: $\kappa_e \propto T$.
In practice, the way to extract $\kappa_e$ is to extrapolate $\kappa/T$ to $T=0$, and thus obtain the purely electronic residual linear term,
$\kappa_0/T$.\cite{Shakeripour2009a,Sutherland2003,Hawthorn2007}
If one can neglect electron-phonon scattering, as one usually can deep in the superconducting state, then the mean free path of phonons as $T \to 0$
is controlled by the sample boundaries. If those boundaries are rough, the scattering is diffuse and the mean free path is constant, such that
the phonon conductivity $\kappa_p \propto T^3$. (Phonons can also be scattered by twin boundaries and grain boundaries.)
If the sample boundaries are smooth, specular reflection yields a temperature-dependent mean free path, and
$\kappa_p \propto T^\alpha$, typically with $2 < \alpha < 3$.\cite{Sutherland2003,Li2008}

%------------------------ FIGURE 5     kappa/T vs T^2 for 6 of the 7 c-axis samples A  ------------------------------------------------

\begin{figure} [t]
\centering
\includegraphics[width=8.5cm]{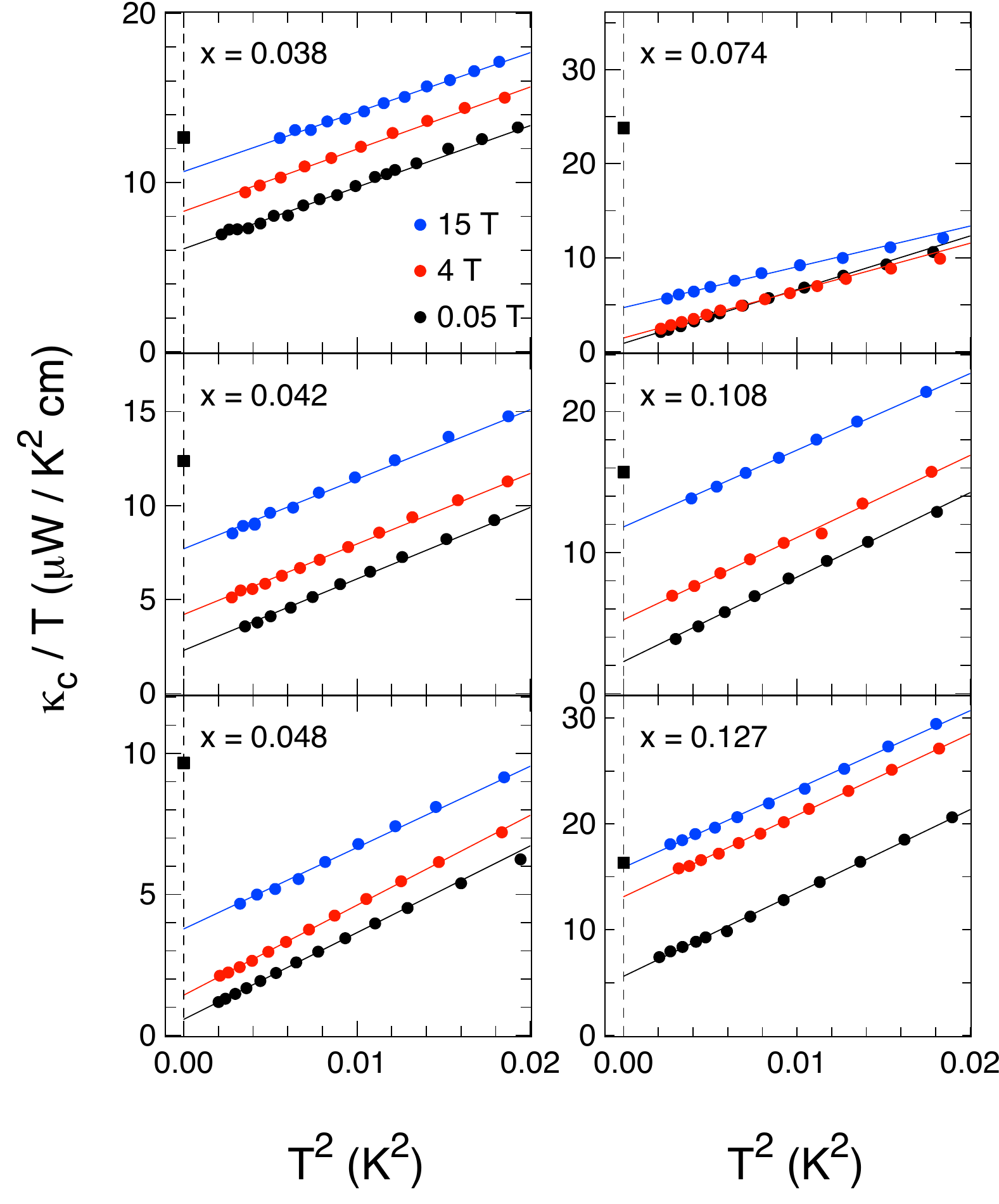}
\caption{
Temperature dependence of the $c$-axis thermal conductivity $\kappa_c$, plotted as $\kappa_c/T$ vs $T^2$, for six samples of Ba(Fe$_{1-x}$Co$_x$)$_2$As$_2$, 
with $x$ as indicated, in a magnetic field $H = 0.05$, 4, and 15~T (data taken at other fields are not shown for clarity).
These are six of the seven samples labelled A in Table~I.
The lines are a linear fit to the data below $T^2 = 0.015$~K$^2$, used to extract the residual linear term $\kappa_{c0}/T$ as the extrapolation of $\kappa_c/T$ to $T=0$.
The values of $\kappa_{c0}/T$ are listed in Table~I for $H=0$, and plotted vs $H$ in Fig.~\ref{kappac0overTvsH}, for all twelve $c$-axis samples.
Solid black squares on the $T=0$ axis (dashed line) give the residual linear term in the normal-state thermal conductivity, $\kappa_{\rm cN}/T$, obtained from the residual resistivity $\rho_{c0}$ of the sample via the Wiedemann-Franz law (see Table~I).   
}
\label{kappacoverTvsT2}
\end{figure}

%--------------------------------------------------------------------------------------------------------------------------------------

%--------------------------- FIGURE 6     k_c0/T vs H for all 12 c-axis samples  ------------------------------------------------------

\begin{figure}[t]
\centering
\includegraphics[width=8.5cm]{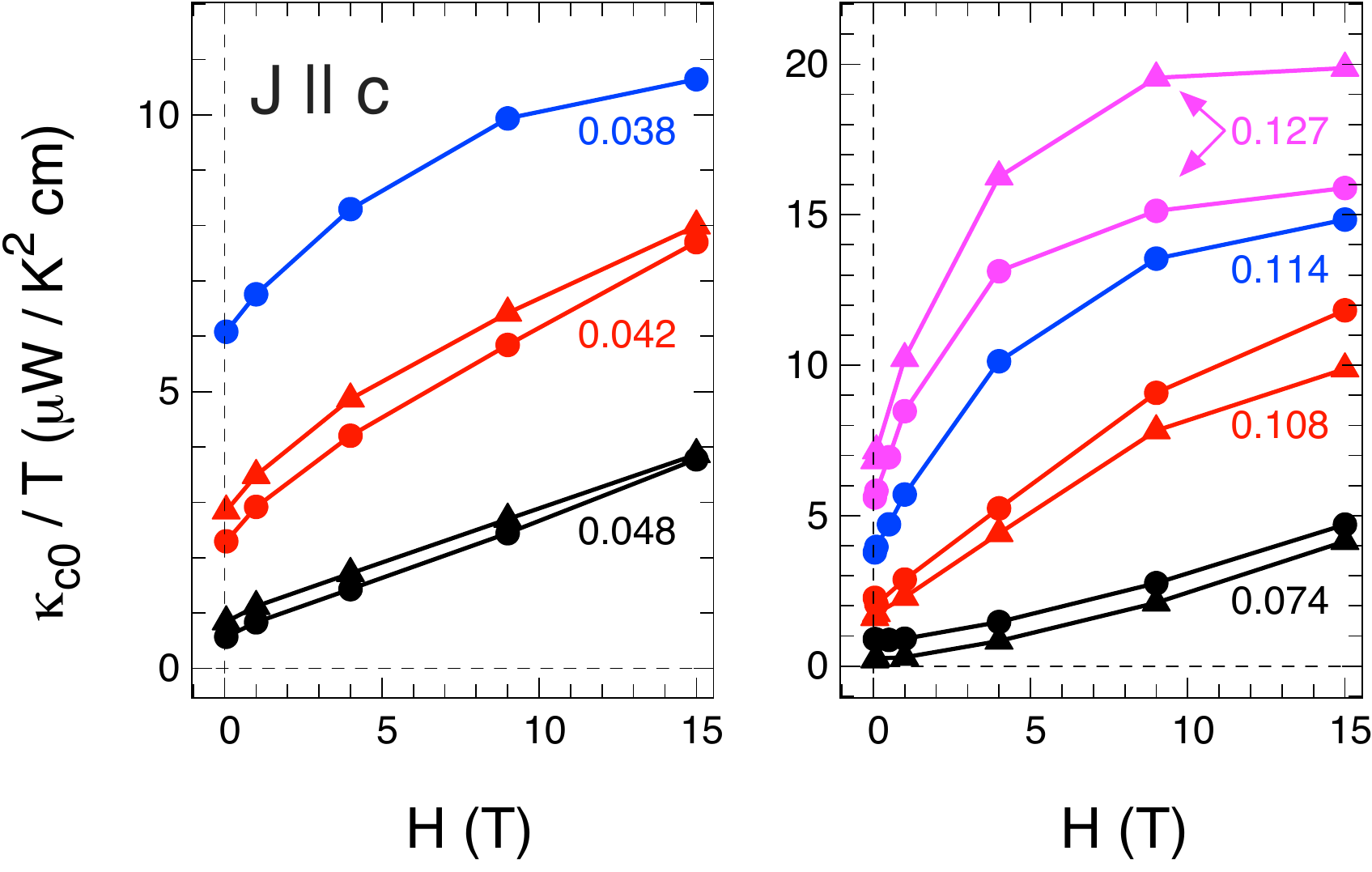}
\caption{
Field dependence of the residual linear term $\kappa_{c0}/T$ in the $c$-axis thermal conductivity of our twelve $c$-axis single crystals of Co-Ba122,
with $x$ as indicated.
Underdoped compositions are shown on the left, overdoped compositions on the right.
For five concentrations, two different crystals with nominally the same $x$ value, labelled A (circles) and B (triangles), were measured (see Table~I).
}
\label{kappac0overTvsH}
\end{figure}

%--------------------------------------------------------------------------------------------------------------------------------------

In Fig.~\ref{kappacoverTvsT2}, we show the thermal conductivity $\kappa_c$ of our $c$-axis samples, plotted as $\kappa_c/T$ vs $T^2$, for magnetic fields from $H=0.05$ to 15~T.
Below $T \simeq 0.15$~K, the curves are linear, consistent with diffuse phonon scattering on the sample boundaries of our $c$-axis samples, which are indeed characterized by rough side surfaces. 
We obtain $\kappa_e/T \equiv \kappa_0/T$ by extrapolating $\kappa/T$ to $T=0$ using a linear fit below $T^2 = 0.015$~K$^2$.
The error bar on this extrapolation is approximately $\pm~0.5~\mu$W/K$^2$~cm, for all $c$-axis samples.
The value of $\kappa_{c0}/T$ thus obtained is plotted as a function of field $H$ in Fig.~\ref{kappac0overTvsH}, for all twelve $c$-axis samples. 
For five concentrations, we have a pair of crystals with nominally the same Co concentration. As can be seen, the two curves in each pair are in good agreement with each other, well within the uncertainty in the geometric factor. 
The zero-field values are listed in Table~I. They range from $\kappa_{c0}/T < 1~\mu$W/K$^2$~cm at $x = 0.048$ and 0.074 to 
$\kappa_{c0}/T \simeq 6~\mu$W/K$^2$~cm at $x = 0.038$ and 0.127.  

% Verification of the Wiedemann-Franz law

The normal-state residual linear term $\kappa_{\rm N}/T$ was estimated using the values of $\rho_0$ through application of the Wiedemann-Franz law. 
The value of $\kappa_{\rm N}/T$ is shown as a solid black square on the $y$ axis of Fig.~\ref{kappacoverTvsT2}. 
For the most heavily overdoped samples, with $x = 0.127$, a magnetic field of 15~T is sufficient 
to reach the normal state, where $\kappa_0/T$ saturates to its normal-state value $\kappa_{\rm N}/T$. 
This allows us to check the Wiedemann-Franz law. 
For sample A, $\kappa_{c0}/T = 16.0 \pm 0.5~\mu$W/K$^2$~cm at $H = 15$~T, while $\kappa_{\rm N}/T = 16.3 \pm 0.8~\mu$W/K$^2$~cm;
for sample B, $\kappa_{c0}/T = 20.0 \pm 0.5~\mu$W/K$^2$~cm at $H = 15$~T, while $\kappa_{\rm N}/T = 21.7 \pm 1.1~\mu$W/K$^2$~cm.  
Within error bars, associated with extrapolations to get $\kappa_{0}/T$ and $\rho_0$, the Wiedemann-Franz law is satisfied in both samples.

\subsection{Heat transport in the $a$ direction} 
 
%------------------------ FIGURE 7     kappa/T vs T for the 6 a-axis samples A  ------------------------------------------------

\begin{figure} [t]
\centering
\includegraphics[width=8.5cm]{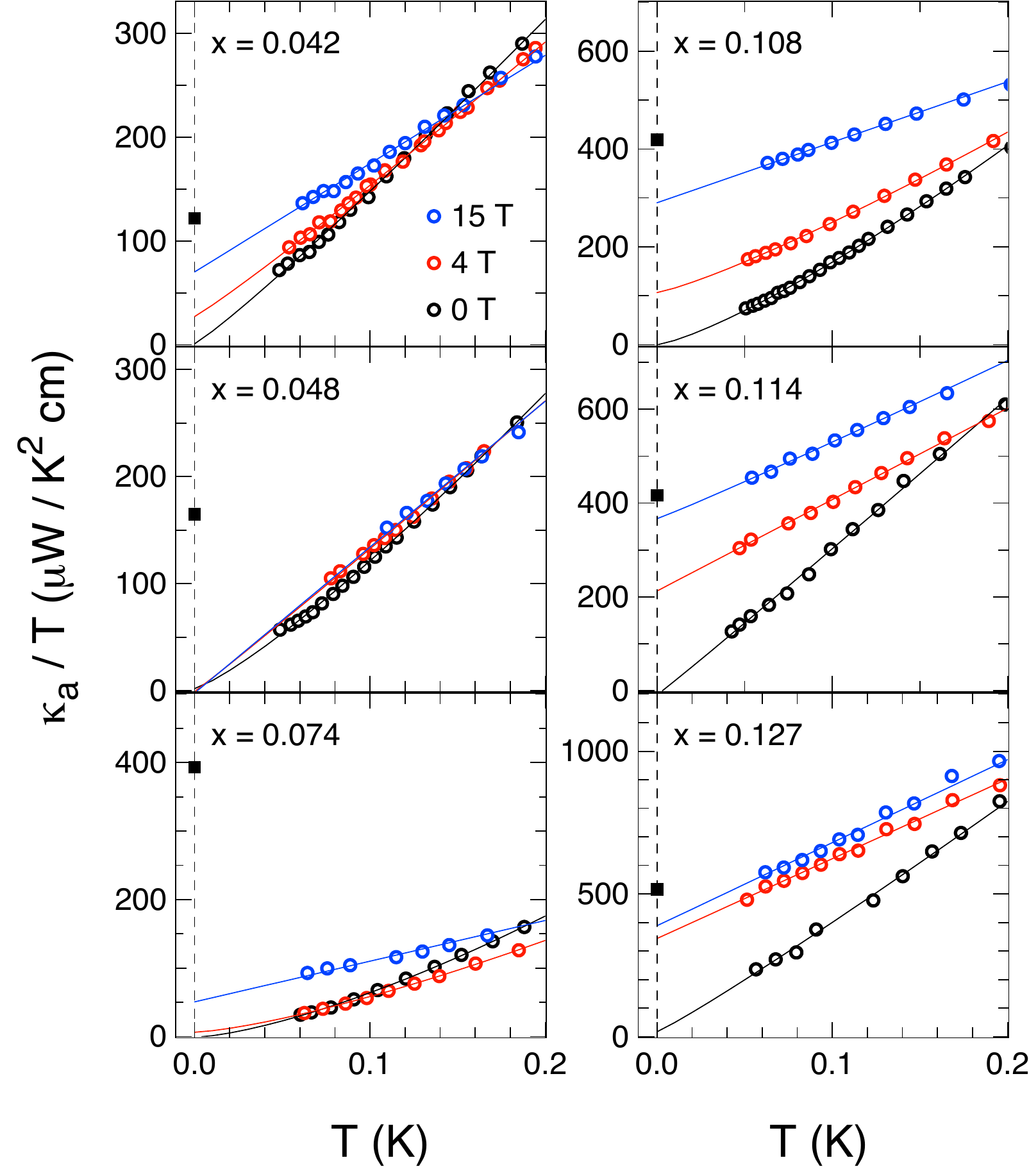}
\caption{
Temperature dependence of the $a$-axis thermal conductivity $\kappa_a$, plotted as $\kappa_a/T$ vs $T$, for six samples of Ba(Fe$_{1-x}$Co$_x$)$_2$As$_2$, 
with $x$ as indicated, in a magnetic field $H = 0$, 4, and 15~T (data taken at other fields are not shown for clarity).
These are the six samples labelled A in Table~II.
%(The curve for $x=0.108$ was published previously~\cite{Tanatar2010}.)
%%
The lines are a power-law fit to the data below $T = 0.3$~K, namely $\kappa/T = a + b T^{\alpha - 1}$.
The fit is used to extrapolate $\kappa_a/T$ to $T=0$ and thus obtain the residual linear term $\kappa_{a0}/T$.
The power $\alpha$ is in the range from 2 to 2.5.
The values of $\kappa_{a0}/T$ are listed in Table~II for $H=0$, and plotted vs $H$ in Fig.~\ref{kappaa0overTvsH}, for all nine $a$-axis samples.
Solid black squares on the $T=0$ axis give the residual linear term in the normal-state thermal conductivity, $\kappa_{\rm aN}/T$, obtained from the residual resistivity $\rho_{a0}$ of the sample via the Wiedemann-Franz law (see Table~II).   
}
\label{kappaaoverTvsT}
\end{figure}

%--------------------------------------------------------------------------------------------------------------------------------------

%--------------------------- FIGURE 8     k_a0/T vs H for all 9 a-axis samples  ------------------------------------------------------

\begin{figure}
\centering
\includegraphics[width=8.5cm]{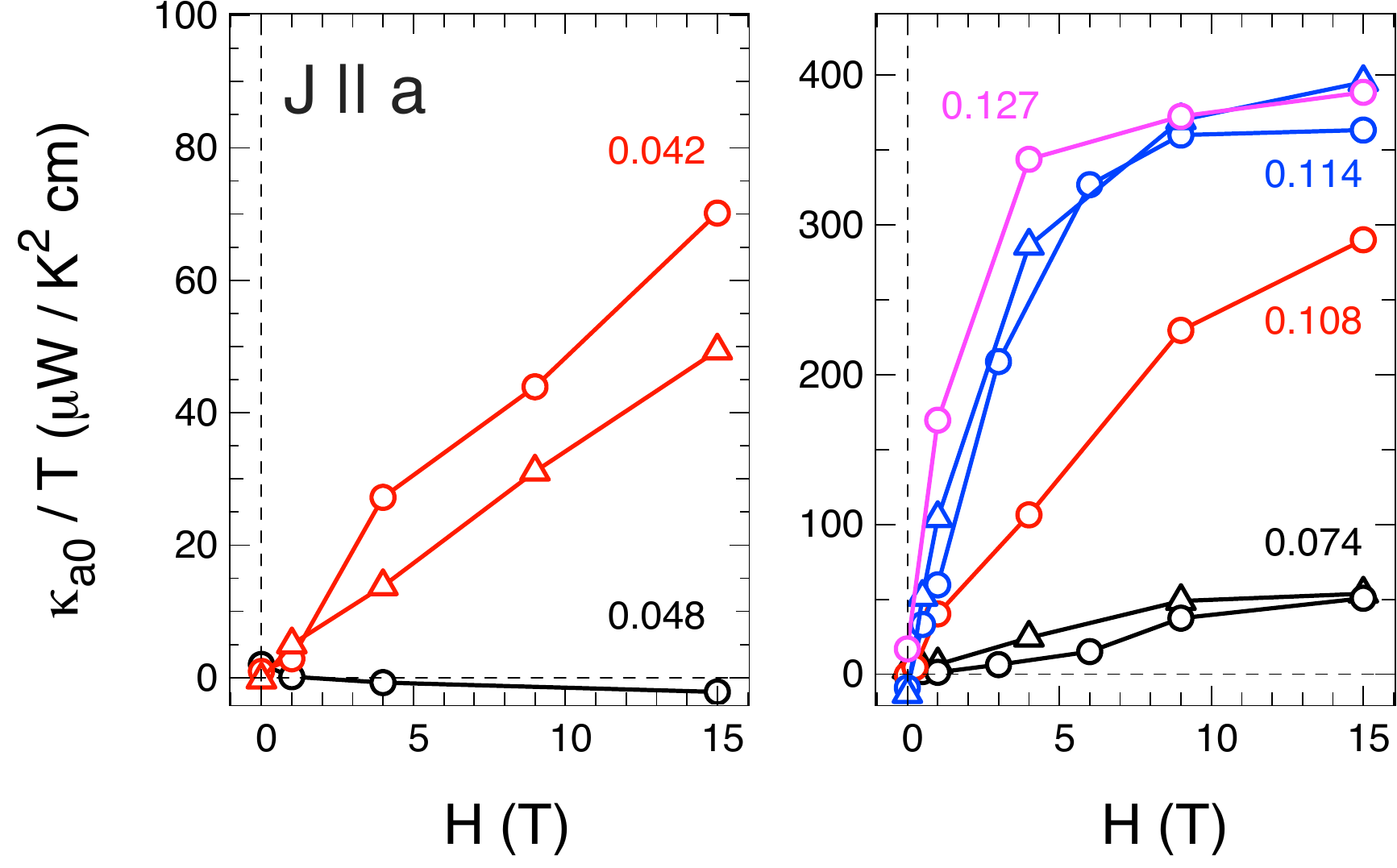}
\caption{
Field dependence of the residual linear term $\kappa_{a0}/T$ in the $a$-axis thermal conductivity of our nine $a$-axis single crystals of Co-Ba122,
with $x$ as indicated.
Underdoped compositions are shown of the left, overdoped compositions on the right.
For three concentrations, two different crystals with nominally the same $x$ value, labelled A (circles) and B (triangles), were measured (see Table~II).
}
\label{kappaa0overTvsH}
\end{figure}

%--------------------------------------------------------------------------------------------------------------------------------------

In Fig.~\ref{kappaaoverTvsT}, we show the thermal conductivity $\kappa_a$ for six of our nine $a$-axis samples, plotted as $\kappa_a/T$ vs $T$, for magnetic fields from $H=0$ to 15~T. 
Unlike in the $c$-axis samples, the phonon conductivity $\kappa_p$ does not obey $\kappa_p/T \propto T^2$ as $T \to 0$. 
Instead, it follows approximately a power law such that $\kappa_p/T \propto T^{\alpha - 1}$, with $2.0 < \alpha < 2.5$. 
These values of $\alpha$ are typical of specular reflection off smooth mirror-like surfaces.\cite{Sutherland2003,Li2008} 
The cleaved surfaces of these Co-Ba122 crystals (normal to the $c$ axis) are indeed mirror-like. 
Previous measurements of in-plane heat transport on K-Ba122,\cite{Luo2009} Co-Ba122,\cite{Dong2010} and Ni-Ba122~\cite{Ding2009} have all obtained $\alpha < 2.7$.
%
%Specular reflection will also lengthen the phonon mean free path. 
%
%This could in part explain the fact that the phonon conductivity is much larger in the basal plane than along the $c$-axis, by an order of magnitude.
%Some of that large anisotropy could also come from an anisotropy of the sound velocity.   

As done previously for other $a$-axis samples,\cite{Tanatar2010} we obtain the residual linear term $\kappa_{a0}/T$ by fitting the data below $T = 0.3$~K to a power-law expression, $\kappa/T = a + bT^{\alpha -1}$, where $a \equiv \kappa_{a0}/T$. 
The error bar on this extrapolation is approximately in the range $\pm~10-20~\mu$W/K$^2$~cm.
(The uncertainty is an order of magnitude larger than for $\kappa_{c0}/T$ because the phonon-related slope is an order of magnitude steeper.)
As found previously over the concentration range $0.048 \leq x \leq 0.114$,\cite{Tanatar2010} we again find $\kappa_{a0}/T \simeq 0$, within error bars, 
now over a wider range: $0.042 \leq x \leq 0.127$.
This is consistent with a separate report that $\kappa_{a0}/T \simeq 0$ in Co-Ba122 at $x = 0.135$.\cite{Dong2010} 

Upon application of a magnetic field, $\kappa_{a0}/T$ increases, as displayed in Fig.~\ref{kappaa0overTvsH} for all nine $a$-axis samples. 
For three concentrations, we have a pair of $a$-axis crystals with nominally the same Co concentration. As can be seen, the two curves in each pair are in good agreement with each other, within the $\pm~20$~\% uncertainty in the geometric factor and the error bar on the extrapolations.

%%%%%%%%%%%%%%%%%%%%%%%%%%%%%%%%%%%%%%%%%%%%%%%%%%%%%%%%%%%%%%%%%%%%%%%%%%%%%%%%%%%%%%%%%%%%%%%%%%%%%%%%%%%%%%%%%%%%%%%%%%%%%%%
%%%%%%%%%%%%%%%%%%%%%%%%%%%%%%%%%%%%%%%%%%%%  DISCUSSION  %%%%%%%%%%%%%%%%%%%%%%%%%%%%%%%%%%%%%%%%%%%%%%%%%%%%%%%%%%%%%%%%%%%%%
%%%%%%%%%%%%%%%%%%%%%%%%%%%%%%%%%%%%%%%%%%%%%%%%%%%%%%%%%%%%%%%%%%%%%%%%%%%%%%%%%%%%%%%%%%%%%%%%%%%%%%%%%%%%%%%%%%%%%%%%%%%%%%%

\section{Discussion} 

\begin{figure} [t]
\centering
\includegraphics[width=8.5cm]{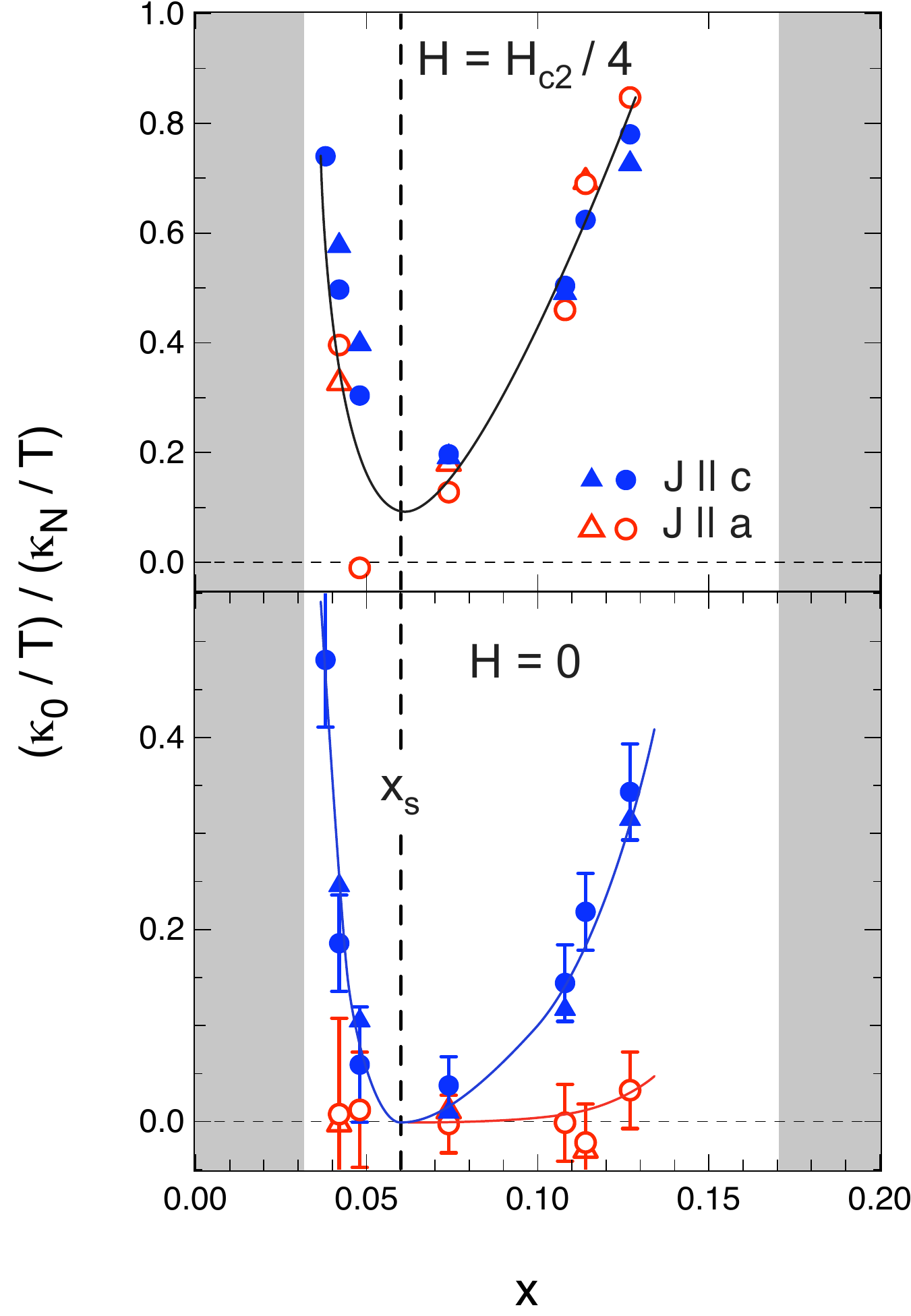}
\caption{
Residual linear term $\kappa_0/T$ of Co-Ba122 normalized by the normal-state value $\kappa_{\rm N}/T$ as a function of Co concentration $x$, at $H=0$ (lower panel) and at 
$H = H_{c2}/4$ (upper panel).
Full blue symbols are for a heat current along the $c$ axis ($J~||~c$; circles for A samples, triangles for B samples in Table~I).
Empty red symbols are for a heat current along the $a$ axis ($J~||~a$; circles for A samples, triangles for B samples in Table~II).
The white interval between the two vertical grey bands at $x < 0.032$ and $x > 0.17$ is the region of superconductivity in the phase diagram (Fig.~\ref{phase-diagram}).
The vertical dashed line at $x = 0.06$ marks the approximate location of the critical doping $x_s$ where the structure at $T=0$ goes from orthorhombic (below) to tetragonal (above)~\cite{Nandi2010} (see Fig.~\ref{phase-diagram}).
%Antiferromagnetic order sets in only slightly below $x_c$, causing a reconstruction of the Fermi surface.
%
Lines through the data points are a guide to the eye.
Error bars on the $H=0$ data are shown for the A samples (circles).
}
\label{kappa0overkappaNvsx}
\end{figure}

The results of our study are summarized in Fig.~\ref{kappa0overkappaNvsx}, where the $\kappa_{0}/T$ values of all 21 samples 
are plotted vs $x$, normalized to their respective normal-state value $\kappa_{\rm N}/T$. 
%This normalization eliminates the uncertainty associated with geometric factors and normalizes out the effect of a change in Fermi surface across the
%orthorhombic/antiferromagnetic boundary (see Fig.~\ref{phasediagram}). 
%

% --------------------------- Figure 9 ---------------------------------------------------------------------------------------

% -----------------------------------------------------------------------------------------------------------------------------

\subsection{Gap nodes}

\subsubsection{Zero magnetic field}

Our central finding is the presence of a substantial residual linear term $\kappa_{0}/T$ in the thermal conductivity of Co-Ba122 in zero field,
for heat transport along the $c$ axis. 
It implies the presence of nodes in the superconducting gap, such that $\Delta({\bf k}) = 0$ for some wavevectors ${\bf k}$ on the 
Fermi surface.\cite{Hirschfeld1988,Graf1996,Durst2000,Mishra2009,Shakeripour2009a} 
Because heat conduction in a given direction is dominated by quasiparticles with ${\bf k}$ vectors along that direction,\cite{Hirschfeld1988,Graf1996}
the fact that $\kappa_{0}/T$ is negligible when heat transport is along the $a$ axis, at all $x$, implies that the nodes are located in regions of the Fermi surface 
that contribute strongly to $c$-axis conduction but very little to in-plane conduction.
%
%
%%%%%%%%%%%%%%%%%%%%%%%%%%%%%%%%%%%     x > x_c  %%%%%%%%%%%%%%%%%%%%%%%%%%%%%%%%%%%%%%%%%%%%%%%%%%%%%%%%%%%%%%%%%%
%
The anisotropy of $\kappa_0/T$ becomes pronounced as $x$ moves away from the critical doping $x_s \simeq 0.06$, in either direction.
For $x > 0.06$, we see that the $a/c$ anisotropy in $\kappa_0/\kappa_{\rm N}$ is at least a factor 10 (see Fig.~\ref{kappa0overkappaNvsx}).
Such a large anisotropy is not expected in a scenario of isotropic pair-breaking,\cite{Kogan2009} and it confirms that the residual linear term seen
in the $c$ direction is due to nodes.

At the highest doping studied here, $x = 0.127$, 
$\kappa_0/\kappa_{\rm N} = 0.34 \pm 0.03$ for $J~||~c$.
(This is for sample A, which has the lowest $T_c$ in the overdoped regime; see Table~I.)
This magnitude is typical of superconductors with a line of nodes in the gap.
In the heavy-fermion superconductor CeIrIn$_5$, with $T_c = 0.4$~K and $H_{c2} \simeq 0.5$~T, 
$\kappa_0/\kappa_{\rm N} \simeq 0.2$.\cite{Shakeripour2007,Shakeripour2009} 
In the ruthenate superconductor Sr$_2$RuO$_4$, with $T_c = 1.5$~K and $H_{c2} = 1.5$~T, 
$\kappa_0/\kappa_{\rm N} \simeq 0.1 - 0.3$~(depending on sample purity).\cite{Suzuki2002} 
In the overdoped cuprate Tl$_2$Ba$_2$CuO$_{6-\delta}$ (Tl-2201), a $d$-wave superconductor with $T_c = 15$~K and $H_{c2} \simeq 7$~T, 
$\kappa_0/\kappa_{\rm N} \simeq 0.35$.\cite{Proust2002} 
In the latter case, because the order parameter is well-known and the Fermi surface is very simple (a single 2D cylinder), it was possible to show that the magnitude of $\kappa_0/T$ agrees quantitatively with the theoretical BCS expression for the residual linear term in a $d$-wave superconductor,\cite{Hawthorn2007} namely 
$\kappa_{0}/T = (k^2_{\rm B}/3 d) (k_{\rm F} v_{\rm F} / S)$,\cite{Graf1996,Durst2000,Shakeripour2009a} where $d$ is the interlayer separation, $k_{\rm F}$ and $v_{\rm F}$ are the Fermi wavevector and velocity at the node, respectively, and $S \equiv \delta \Delta / \delta k$ is the slope of the gap at the node.
(For a $d$-wave gap with $\Delta(k)=\Delta_0$cos($2\phi$), $S = 2\Delta_0$.)  

% --------------------------- FIGURE 10 -----------------------------------------------------------------------------------------
%
\begin{figure}
\centering
\includegraphics[width=8.5cm]{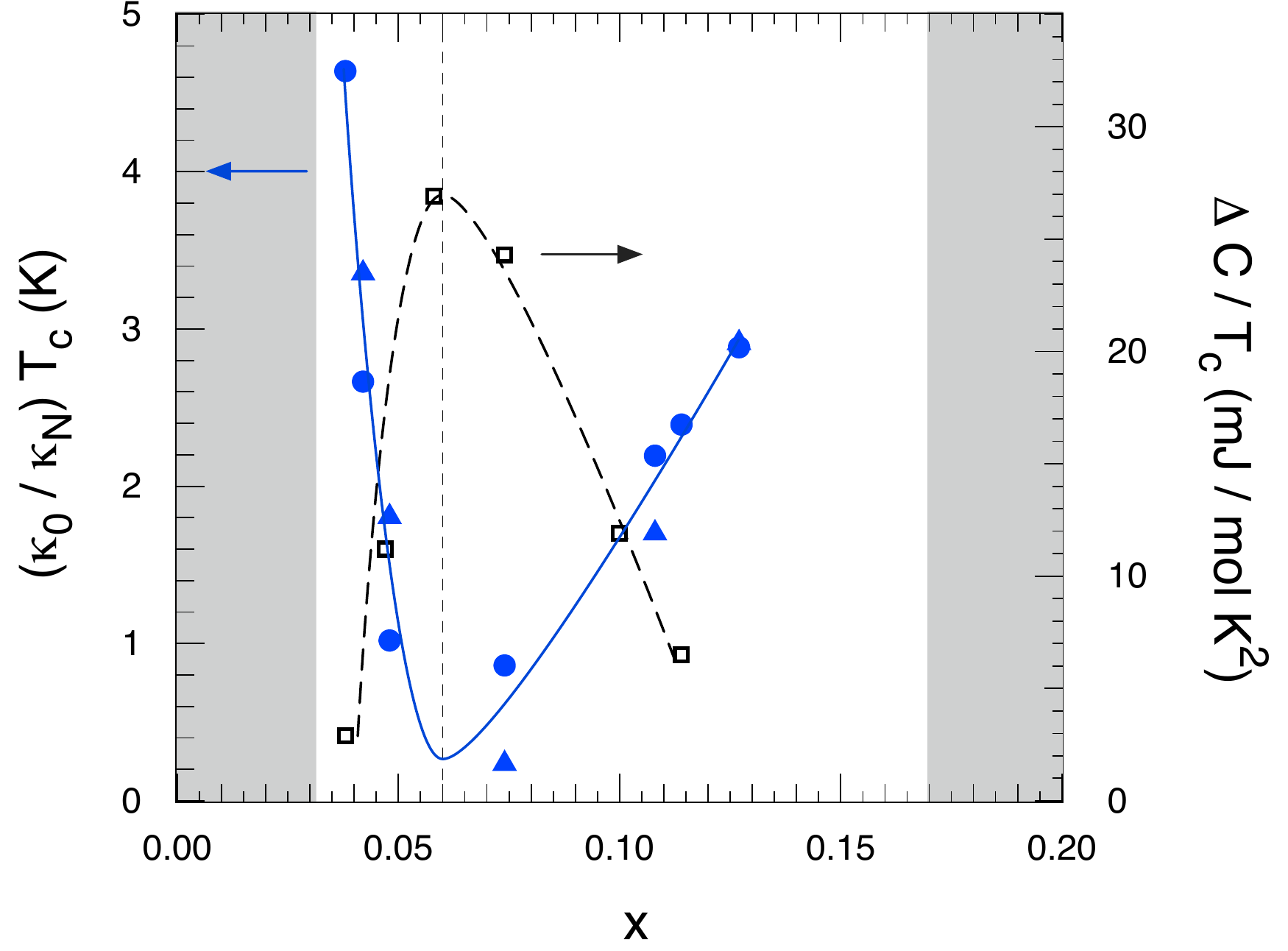}
\caption{
Comparison of heat transport by nodal quasiparticles and the jump in heat capacity at the superconducting transition, in Co-Ba122 as a function of Co concentration $x$. 
Heat transport is measured as the zero-field residual linear term in the thermal conductivity along the $c$ axis, $\kappa_{c0}/T$, normalized by the corresponding normal-state conductivity at $T \to 0$, $\kappa_{\rm cN}/T$, multiplied by $T_c$. 
The heat capacity jump $\Delta C$ is divided by $T_c$ (from Ref.~\onlinecite{Budko2009}).  
The vertical dashed line marks the location of $x_s$.
Other lines are a guide to the eye.
}
\label{comparisonwithDeltaCoverTc}
\end{figure}
%
% ---------------------------------------------------------------------------------------------------------------------------------
%

If the line of nodes in the gap is imposed by the symmetry of the order parameter, as in a $d$-wave state, then
$\kappa_{0}/T$ is universal, {\it i.e.} independent of the impurity scattering rate $\Gamma_0$, in the clean limit $\hbar \Gamma_0 << \Delta_0$.\cite{Graf1996,Durst2000}
Such universal transport was demonstrated experimentally for CeIrIn$_5$,\cite{Shakeripour2009} Sr$_2$RuO$_4$,\cite{Suzuki2002} and the cuprates YBa$_2$Cu$_3$O$_7$~\cite{Taillefer1997} and Bi$_2$Sr$_2$CaCu$_2$O$_8$.\cite{Nakamae2001}
As a fraction of the normal-state conductivity, one then gets $(\kappa_0/T)/(\kappa_{\rm N}/T) \equiv \kappa_0/\kappa_{\rm N} \propto \hbar \Gamma_0 / S$.\cite{Graf1996}
However, if the nodes are not imposed by symmetry, but are `accidental', as in an `extended-$s$-wave' state, they still cause a non-zero residual linear term, with $\kappa_0/T \propto 1/S$, but $\kappa_0/T$ is no longer universal, because $S$ depends on the scattering rate $\Gamma_0$.\cite{Mishra2009}

In Fig.~\ref{kappa0overkappaNvsx}, we see that $\kappa_{c0}/\kappa_{\rm cN}$ exhibits
a striking U-shaped dependence on Co concentration $x$,
with $\kappa_{c0}/\kappa_{\rm cN} \to 0$ as $x \to x_s$.
Just above $x_s$, at $x = 0.074$, 
$\kappa_{c0}/T = 0.2 \pm 0.5~\mu$W/K$^2$~cm.
(This is for sample B, which is closest to $x_s$, as it has the highest $T_c$; see Table~I.)
This is equal to zero within error bars, indicating that there are no nodes in the gap at this concentration, 
as also inferred from the field dependence (see below).
If the nodes can be removed simply by changing $x$, then these nodes must be accidental,
not imposed by symmetry.

Given that the change in $\kappa_0/\kappa_{\rm N}$ with $x$ on the overdoped side is due to a change in $\kappa_0/T$ and not a change in $\kappa_{\rm N}/T$ (since
$\rho_0$ is independent of $x$, within error bars), we attribute the dramatic rise in $\kappa_0/\kappa_{\rm N}$ 
from $x = 0.074$ to $x = 0.127$ to a decrease of the slope $S$ with increasing $x$.
Part of this decrease must be due to a drop in the overall strength of superconductivity, as measured by the decreasing $T_c$. 
We can factor out that effect by multiplying $\kappa_0/\kappa_{\rm N}$ by $T_c$, as shown in Fig.~\ref{comparisonwithDeltaCoverTc}. 
We see that $\kappa_0/\kappa_{\rm N} \times T_c$ vs $x$ is far from constant, as it would be if the decrease of $\Delta({\bf k})$ vs $x$
was uniform, independent of ${\bf k}$.
In a $d$-wave superconductor, for example, $S$ would typically scale with the gap maximum $\Delta_0$, which itself would scale with $T_c$, giving a constant
product $\kappa_0/\kappa_{\rm N} \times T_c$ (for a constant $\Gamma_0$).
By contrast, in Co-Ba122 the slope of the gap at the nodes decreases faster than that part of the gap structure which controls $T_c$. 
In other words, $\Delta({\bf k})$ must be acquiring a stronger and stronger ${\bf k}$ dependence, or modulation, with increasing $x$.

%%%%%%%%%%%%%%%%%%%%%%%%%%%%%%%%%%%     x < x_c  %%%%%%%%%%%%%%%%%%%%%%%%%%%%%%%%%%%%%%%%%%%%%%%%%%%%%%%%%%%%%%%%%%

In the underdoped regime, for samples with $x = 0.048$ and lower, the metal is antiferromagnetic~\cite{Fernandes2010}
and its Fermi surface is reconstructed by the antiferromagnetic order.
Nevertheless, a residual linear term $\kappa_{c0}/T$ is still observed at $H=0$~(see Fig.~\ref{kappa0overkappaNvsx}).
At $x = 0.038$, it is even larger than at $x = 0.127$, namely $\kappa_0/\kappa_N = 0.48 \pm~0.04$~(see Table~I).
This implies that nodes are present in the superconducting gap inside the region of co-existing antiferromagnetic order. 
The fact that $\kappa_{0}/T$ is again strongly anisotropic (see Fig.~\ref{kappa0overkappaNvsx}) 
means that those nodes are still located in regions of the Fermi surface
that contribute strongly to $c$-axis conduction and little to $a$-axis conduction.
The fact that the nodes survive the Fermi-surface reconstruction is consistent with their location in regions with strong 3D character, 
since the spin-density wave gaps the nested portions of the Fermi surface, which are typically those with strong 2D character.
(It should be emphasized that the mechanisms responsible for the drop in $T_c$ and the rise in $\kappa_0/\kappa_{\rm N}$ are likely to be different
above and below optimal doping.)

\subsubsection{Field dependence}

\begin{figure} [t]
\centering
\includegraphics[width=8.5cm]{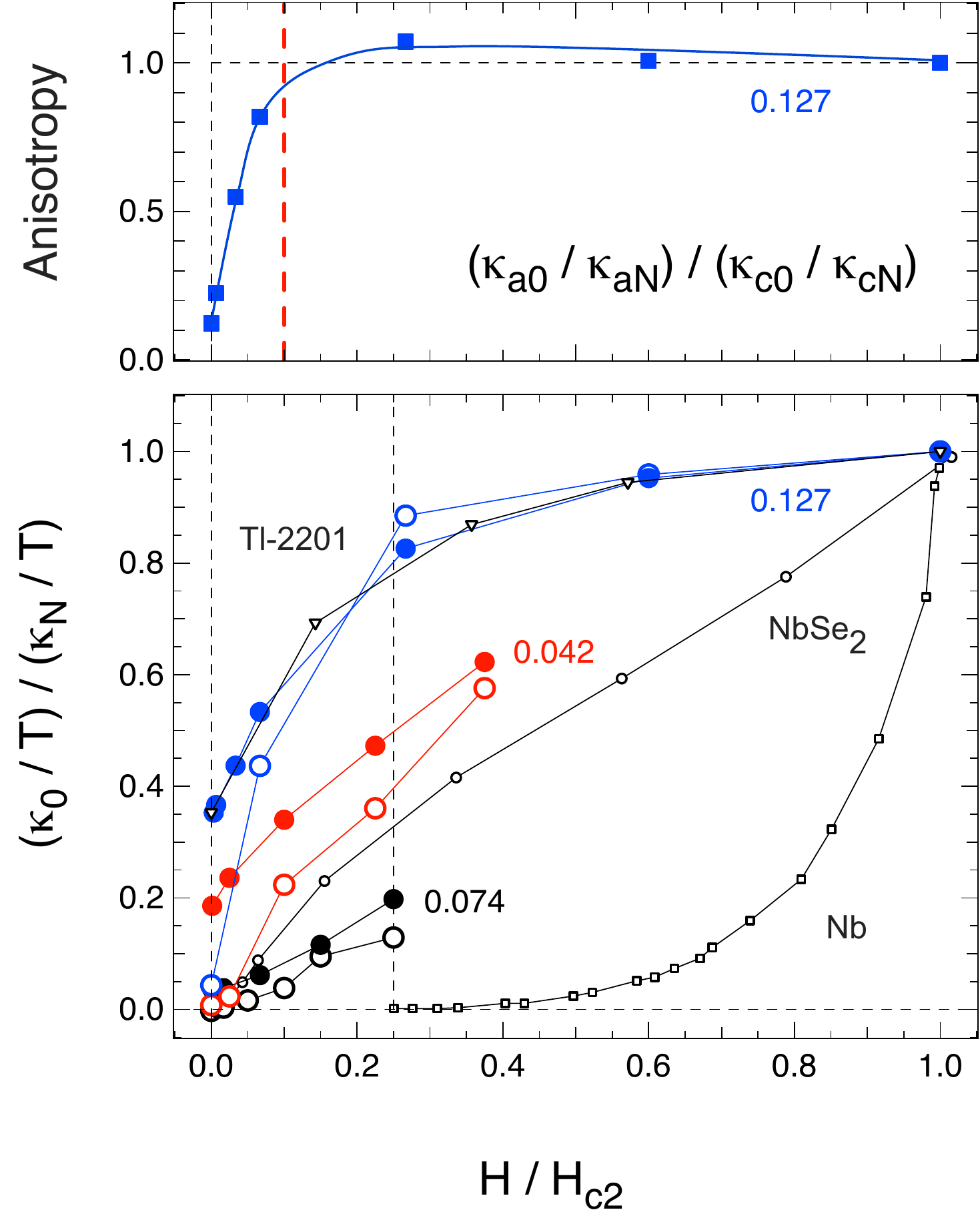}
\caption{ 
Bottom panel: 
Residual linear term $\kappa_0/T$ of Co-Ba122 normalized by the normal-state value $\kappa_{\rm N}/T$ as a function of magnetic field $H$, 
plotted as $(\kappa_0/T)/(\kappa_{\rm N}/T) \equiv \kappa_0/\kappa_{\rm N}$ vs $H/H_{c2}$ for three representative Co concentrations, as indicated: 
underdoped ($x=0.042$; red), 
slightly overdoped ($x=0.074$; black), 
and strongly overdoped ($x=0.127$; blue).
$\kappa_{\rm N}/T$ is obtained from the Wiedemann-Franz law (see text and Tables~I and II), except for the samples with $x = 0.127$ where we use the value of
$\kappa_0/T$ measured at $H=15$~T, since $H_{c2} = 15$~T at that concentration.
%The upper critical field $H_{c2}$ in the $T \to 0$ limit is taken from the literature, as given in Tables~I and II.
%
Full circles are for a heat current along the $c$ axis ($J~||~c$; data from three A samples - see Table~I).
Empty circles are for a heat current along the $a$ axis ($J~||~a$; data from three A samples - see Table~II).
The vertical dashed line marks $H = H_{c2}/4$; the value of $\kappa_0/\kappa_{\rm N}$ at $H_{c2}/4$ is plotted in the top panel of Fig.~\ref{kappa0overkappaNvsx}, 
for all $x$.
We also reproduce corresponding data for the $d$-wave superconductor Tl-2201~(from Ref.~\onlinecite{Proust2002}), 
the isotropic $s$-wave superconductor Nb, and
the multi-band $s$-wave superconductor NbSe$_2$ (from Ref.~\onlinecite{Boaknin2003}).
Top panel: Anisotropy of the normalized residual linear term $\kappa_0/\kappa_{\rm N}$ at $x=0.127$. 
The red dashed line at $H_{c2}/10$ marks roughly the field beyond which $\kappa_0/\kappa_{\rm N}$ becomes isotropic. 
}
\label{kappa0overkappaNvsHoverHc2}
\end{figure}

The effect of a magnetic field on $\kappa_0/T$ reveals how easy it is to excite quasiparticles at $T=0$.\cite{Shakeripour2009a,Mishra2009,Kubert1998}
For a gap with nodes, the rise in $\kappa_0/T$ with $H$ is very fast, because delocalized quasiparticles exist outside the vortices,\cite{Kubert1998}
as shown for the $d$-wave superconductor Tl-2201 in Fig.~\ref{kappa0overkappaNvsHoverHc2}.
For a full gap without nodes or deep minima, such as in the $s$-wave superconductor Nb, 
the rise in $\kappa_0/T$ vs $H$ is exponentially slow (see Fig.~\ref{kappa0overkappaNvsHoverHc2}), 
because it relies on tunneling between quasiparticle states localized on adjacent vortices.
For Co-Ba122 at $x = 0.127$, $\kappa_{c0}/\kappa_{\rm N}$ is seen to track the $d$-wave data all the way from $H=0$ to $H = H_{c2}$.
This nicely confirms the presence of nodes in the gap structure of overdoped Co-Ba122 that dominate the transport along the $c$ axis.

% --------------------------- Figure 11 ---------------------------------------------------------------------------------------

% -----------------------------------------------------------------------------------------------------------------------------

By contrast, at $x = 0.074$, the initial rise in $\kappa_{c0}/T$ vs $H$ has the positive (upwards) curvature typical of a nodeless gap, for both samples A 
and B (see Fig.~\ref{kappac0overTvsH}).
The rise at low $H$ is faster than in a simple $s$-wave superconductor like Nb (Fig.~\ref{kappa0overkappaNvsHoverHc2}), 
either because of a $k$-dependence of the gap or because of a multi-band 
variation of the gap amplitude, or both. A multi-band variation is what causes the fast initial rise in $\kappa_0/T$ vs $H$ (with positive curvature) in NbSe$_2$~\cite{Boaknin2003} (see Fig.~\ref{kappa0overkappaNvsHoverHc2}).
This $H$ dependence strongly suggests that there are no nodes in the gap of Co-Ba122 at $x = 0.074$, as inferred above from the negligible value of $\kappa_{a0}/T$.

\subsection{Gap minima}

We saw that nodes in the gap have two general and related signatures in the thermal conductivity:\cite{Shakeripour2009a} 
1) a finite residual linear term $\kappa_0/T$ in zero field, and 
2) a fast initial rise in $\kappa_0/T$ with $H$.
Both signatures are clearly observed in Co-Ba122 at $x = 0.127$ for $J~||~c$.
For $J~||~a$, however, the situation is quite different.
Indeed, $\kappa_{a0}/T$ is negligible at $H=0$, for all $x$, as also found in previous measurements of $\kappa_a$ on underdoped
K-Ba122,\cite{Luo2009} optimally-doped Ni-Ba122,\cite{Ding2009} and overdoped Co-Ba122.\cite{Dong2010}

Consequently, the fast initial rise in $\kappa_0/T$ with $H$ for $J~||~a$, seen in Fig.~\ref{kappa0overkappaNvsHoverHc2}, is not due to nodes but rather to the presence of deep minima in the gap, in regions of the Fermi surface that contribute significantly to in-plane conduction, as previously reported.\cite{Tanatar2010} 
%
%The behaviour of $\kappa_a/T$ in Co-Ba122 is similar to what is observed in the borocarbide superconductor LuNi$_2$B$_2$C~\cite{Boaknin2002}.
%
In the top panel of Fig.~\ref{kappa0overkappaNvsx}, we show the normalized residual linear term $\kappa_{0}/\kappa_{\rm N}$ measured 
at $H = H_{c2}/4$. 
We see that in the overdoped regime the $\kappa_{0}/\kappa_{\rm N}$ values are the same for both current directions, at all $x$.
In other words, whereas $\kappa_{0}/\kappa_{\rm N}$ is very anisotropic at $H=0$, it is essentially isotropic at $H > H_{c2}/10$,
as shown for $x=0.127$ in the top panel of Fig.~\ref{kappa0overkappaNvsHoverHc2}.
But quasiparticle transport for $J~||~c$ is due to nodal excitations, whereas quasiparticle transport for $J~||~a$ comes from field-induced excitations 
across a minimum gap.
In a single-band model, say with a single ellipsoidal Fermi surface, this contrast between zero-field anisotropy and finite-field isotropy
can only be described by invoking two unrelated features in the gap structure $\Delta({\bf k})$: 
nodes along the $c$ axis and deep minima in the basal plane.
However, the fact that $\kappa_0/\kappa_{\rm N}$ remains isotropic at all $x$ (for $H = H_{c2}/4$) strongly suggests that nodes and minima are in fact intimately related.
We therefore propose that they both come from the same tendency of the gap function $\Delta({\bf k})$ to develop a strong modulation as a function of ${\bf k}$, which causes a deep minimum on one Fermi surface and an even deeper minimum on another Fermi surface, where the gap would actually go to (or through) zero.
In other words, instead of invoking two unrelated features of the gap structure on a single Fermi surface, we invoke a single property of the gap structure
which leads to two related manifestations on separate Fermi surfaces.

\subsection{Two simple models for the gap structure}

\begin{figure} [t]
\centering
\includegraphics[width=8cm]{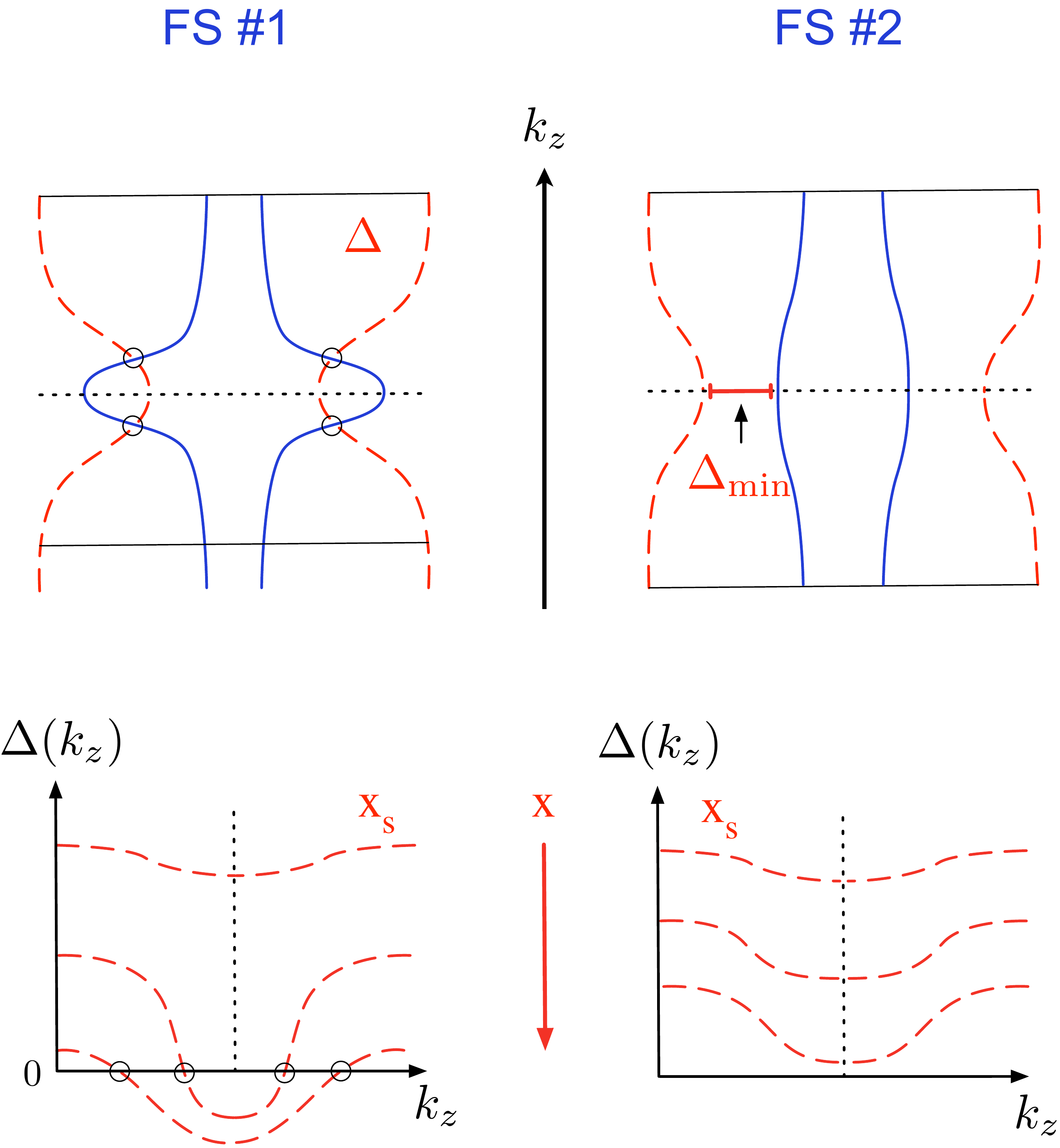}
\caption{ 
First model for the gap structure.
Simplified two-band model of the Fermi surface of Co-Ba122 (blue solid line), shown in the $a-c$ plane, with $k_z~||~c$ in the vertical direction. 
Although in reality the Fermi surface of Co-Ba122 consists of at least four sheets, in our model we reduce it to two sheets:
a sheet with strong 3D character (FS~$\#$ 1; top left) and a sheet with quasi-2D character (FS~$\#$2; top right).
The superconducting gap $\Delta({\bf k})$ on both Fermi surfaces (red dashed line) varies strongly as a function of $k_z$, as shown on the bottom for three representative concentrations $x$ from $x_s$ upwards.
On the 3D Fermi surface ($\#$1), the gap modulation is such that it extends to negative values, producing nodes (black circles) at certain points. 
On the 2D Fermi surface ($\#$2), the gap modulation is strong enough to cause a deep minimum (where $\Delta = \Delta_{\rm min}$), but not nodes.
At $x = x_s$, the gap minima are shallow and there are no nodes on either Fermi surface. 
With increasing $x$ (decreasing $T_c$), the modulation increases, the minima deepen and the nodes appear.
A further increase in $x$ (beyond the maximal concentration in this study) could eventually also yield nodes on FS~$\#$2.
}
\label{FSmodel}
\end{figure}

\begin{figure} [t]
\centering
\includegraphics[width=8.5cm]{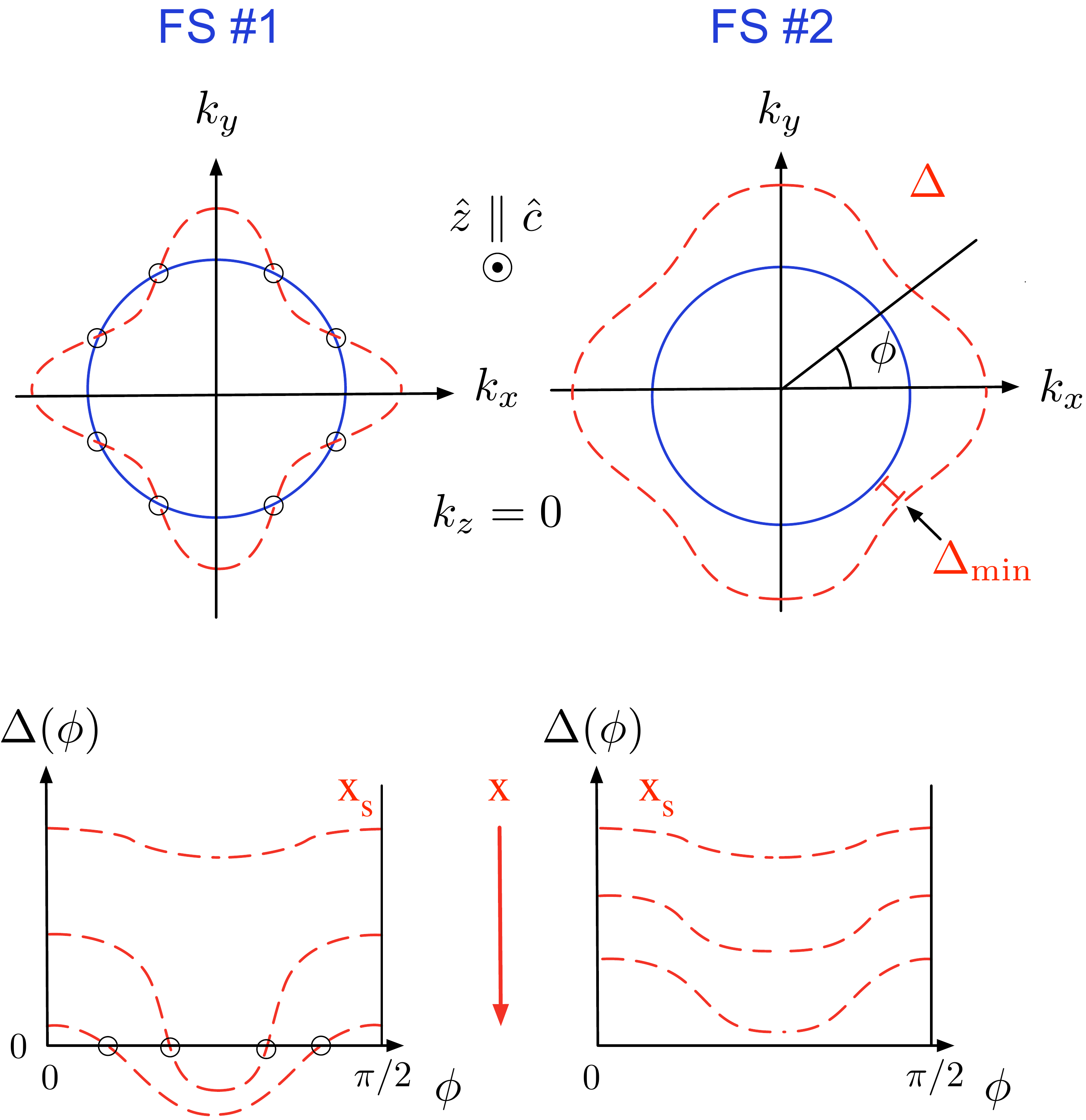}
\caption{ 
Second model for the gap structure.
Same as in Fig.~\ref{FSmodel}, except that the modulation of the 
superconducting gap $\Delta({\bf k})$ is now a function of the azimuthal angle $\phi$ in the basal plane.
}
\label{FSmodel-2}
\end{figure}

For the purpose of illustration, we consider a simplified two-band model for the Fermi surface,  
whereby one surface has strong 3D character and the other has quasi-2D character, as sketched in Fig.~\ref{FSmodel}. 
The 3D Fermi surface can either be open along the $c$ axis, as drawn in Fig.~\ref{FSmodel} and suggested by some ARPES data,\cite{Malaeb2009} 
or closed, as suggested by some band structure calculations.\cite{Mazin2009}
The 3D Fermi surface is responsible for most of the $c$-axis conduction and the 2D surface for most of the $a$-axis conduction
(recall that in the $T=0$ limit $\kappa_{\rm aN}/\kappa_{\rm cN} = \rho_{\rm cN}/\rho_{\rm aN} \simeq 20$). 
Note that in reality the Fermi surface of Co-Ba122 contains at least four separate sheets;\cite{Mazin2009,Graser2010} 
our model requires that at least one of these has strong 3D character and it treats all others in terms of a single Fermi surface, the second quasi-2D sheet.
We then propose that the gap $\Delta({\bf k})$ varies strongly as a function of ${\bf k}$, on both Fermi surface sheets.
There are two basic scenarios: a gap modulation as a function of $k_z$, illustrated in Fig.~\ref{FSmodel}, 
or a gap modulation as a function of the azimuthal angle $\phi$ in the basal plane, illustrated in Fig.~\ref{FSmodel-2}.
The strong modulation extends to negative values on the 3D Fermi surface, thereby producing nodes where $\Delta({\bf k})=0$, 
whereas it only produces a deep minimum (where $\Delta = \Delta_{\rm min}$) on the 2D Fermi surface (at least in the range of concentrations covered here).
In the first scenario (Fig.~\ref{FSmodel}), the lines of nodes are horizontal circular loops in a plane normal to the $c$ axis;
in the second scenario (Fig.~\ref{FSmodel-2}), they are vertical lines along the $c$ axis.

Both versions of the model explain the isotropy at $H_{c2}/4$ and the anisotropy at $H=0$.
The isotropy of $\kappa_0/\kappa_{\rm N}$ follows fundamentally from having a similar ${\bf k}$ modulation of the gap on both Fermi surfaces. 
When the field is large enough to excite quasiparticles across the minimum gap on the 2D Fermi surface, 
quasiparticle transport from both Fermi surfaces will be similar, explaining the rapid and isotropic rise in $\kappa_0/T$ with $H$.
By contrast, at $H=0$ no quasiparticles are excited on the 2D Fermi surface at $T=0$ (since $k_{\rm B}T<< \Delta_{\rm min}$), 
whereas nodal quasiparticles are always present on the 3D surface.
This explains the large anisotropy of $\kappa_0/\kappa_{\rm N}$ at $H=0$.
Note that this anisotropy is not governed by the anisotropy of the gap itself, {\it i.e.} by the direction of the gap modulation,
but rather by the fact that the nodes lie on the 3D Fermi surface.
(Whether horizontal or vertical line nodes are more consistent with our data depends on details of the real Fermi surface of Co-Ba122.)
The nodal quasiparticles on the 3D sheet must also contribute to $a$-axis conduction.
Assuming that for the 3D Fermi surface $\kappa_{a0}/T \simeq \kappa_{c0}/T$ at $H=0$, 
we should detect a residual linear term $\kappa_{a0}/T \simeq 6~\mu$W/K$^2$~cm in the $a$-axis sample with $x = 0.127$, for example.
This is indeed consistent, within error bars, with the value we extrapolate for the $a$-axis data at $x=0.127$ (Fig.~\ref{kappaaoverTvsT}), namely 
$\kappa_{a0}/T = 17~\pm~20~\mu$W/K$^2$~cm (Table~II).

% --------------------------- Figure 12 ---------------------------------------------------------------------------------------

% -----------------------------------------------------------------------------------------------------------------------------

% --------------------------- Figure 13 ---------------------------------------------------------------------------------------

% -----------------------------------------------------------------------------------------------------------------------------

In both versions of our model for the gap structure, the U-shaped $x$ dependence of $\kappa_0/\kappa_{\rm N}$ is attributed to an increase in the modulation of the gap as $x$ moves away from $x_s$, as illustrated in Figs.~\ref{FSmodel} and ~\ref{FSmodel-2}.
The fact that the U-shaped curves in Fig.~\ref{kappa0overkappaNvsx} have their minimum where the (inverted U-shaped) $T_c$ vs $x$ curve has its maximum 
points to a reverse correlation between $T_c$ and gap modulation. Modulation is a sign of weakness.
The presence of nodes in the gap may then be an indicator that pairing conditions are less than optimal.   
%
%There seems to be a correlation between the decrease in $T_c$ and the increase in the gap modulation. Because the nodes are located on the Fermi surface region with %strongest 3D character, {\it i.e.} where $k_z$ dispersion is strongest, it is tempting to relate $k_z$ dispersion of Fermi surface and $k_z$ modulation of $\Delta(k)$.
%In other words, 3D character may cause a softening of $\Delta(k)$ at certain $k$ vectors and a concomittant weakening of superconductivity.
%The degree of $c$-axis dispersion in the Fermi surface may then be an important factor in what controls the maximal value of $T_c$ in various materials of the pnictide %family.

It is possible that at high enough $x$ in the overdoped regime $\Delta_{\rm min}$, the minimum value of the gap on the quasi-2D Fermi surface, 
goes to zero, so that nodes appear on that Fermi surface as well.
This would immediately cause $\kappa_{a0}/T$ to become sizable.
It is conceivable that the large value of $\kappa_{a0}/T$ measured in undoped KFe$_2$As$_2$,\cite{Dong2010a}
which can be viewed as the strong doping limit of K-Ba122, is the result of a gap modulation
so strong that it goes to (through) zero on all Fermi surfaces.  

A pronounced modulation of $\Delta({\bf k})$ should manifest itself in a number of physical properties.
For example, in an $s$-wave superconductor, a variation of the gap magnitude over the Fermi surface, whether from band to band as in MgB$_2$,\cite{Bouquet2002} or from ${\bf k}$ dependence (anisotropy) as in Zn, \cite{Gubser1973} leads to a suppressed ratio of specific heat jump $\Delta C$ at the transition to $T_c$. 
The pronounced gap modulation and anisotropy revealed by the thermal conductivity could therefore account for the dramatic variation of $\Delta C / T_c$ measured in Co-Ba122 vs $x$,\cite{Budko2009} reproduced in Fig.~\ref{comparisonwithDeltaCoverTc}. 
$\Delta C / T_c$ is seen to be maximal where $\kappa_0/\kappa_{\rm N}$ is minimal, {\it i.e.} where the gap modulation is weakest, and it drops just as rapidly with a change in $x$ as $\kappa_0/\kappa_{\rm N}$ rises.

\subsection{Theoretical calculations}

The two-band picture suggested by our thermal conductivity data is reminiscent of the proximity scenario proposed for the three-band quasi-2D $p$-wave 
superconductor Sr$_2$RuO$_4$,\cite{Zhitomirsky2001a} where superconductivity originates on one band, the most 2D one, and is induced by proximity on the other two bands.  
%which have stronger $k_z$ dispersion. 
This $k$-space promixity effect is such that a $k_z$ modulation of the induced gap produces horizontal line nodes on the latter two Fermi surfaces,\cite{Zhitomirsky2001a}
in analogy with the horizontal-line scenario of Fig.~\ref{FSmodel}.
A proximity scenario of this sort was in fact proposed for the pnictides,\cite{Laad2009} predicting $c$-axis nodes in the superconducting gap.
The effect on the superconducting gap structure of including the $k_z$ dispersion of the Fermi surface in BaFe$_2$As$_2$ was recently calculated within a spin-fluctuation pairing mechanism on a 3D multi-orbital Fermi surface.\cite{Graser2010} 
A strong modulation of the gap $\Delta({\bf k})$ as a function of both $k_z$ and $\phi$ is obtained which can indeed, for some parameters, lead to accidental nodes.
%
%These calculations therefore show that it is reasonable to attribute the gap nodes detected in $\kappa_{c}$ to a $k_z$ modulation of $\Delta({\bf k})$, 
%as proposed here in our simple two-band model.

The thermal conductivity of pnictides was calculated in a 2D two-band model for the case of an extended-$s$-wave gap (of $A_{1g}$ symmetry).\cite{Mishra2009}  
These calculations show that the presence of deep minima in the gap, in this case as a function of $\phi$, 
can account for the rapid initial rise observed in $\kappa_{a0}/T$ vs $H$, starting from $\kappa_{a0}/T = 0$ at $H = 0$.
It seems clear that calculations for a gap whose deep minima occur instead as a function of $k_z$ would yield similar results.
It will be interesting to see what calculations of the thermal conductivity give when applied to the 3D model of Ref.~\onlinecite{Graser2010}, or indeed to the simple two-band models proposed here (in Figs.~\ref{FSmodel} and~\ref{FSmodel-2}).

\section{Conclusions}

In summary, our measurements of the thermal conductivity in the iron-arsenide superconductor Ba(Fe$_{1-x}$Co$_x$)$_2$As$_2$ show unambiguously that the gap 
$\Delta({\bf k})$ has nodes.
These nodes are present in both the overdoped and the underdoped regions of the phase diagram, implying that they survive the Fermi-surface reconstruction provoked by the antiferromagnetic order in the underdoped region.
The nodes are located in regions of the Fermi surface that dominate $c$-axis conduction and contribute very little to in-plane conductivity.
The fact that the strongly anisotropic quasiparticle transport at $H=0$ becomes isotropic in a magnetic field $H = H_{c2}/4$ shows that there must be a deep minimum in the gap in regions of the Fermi surface that dominate in-plane transport.
These two features - nodes on 3D regions and minima on 2D regions of the Fermi surface - point to a strong modulation of the gap as a function of 
${\bf k}$. 
This modulation of $\Delta({\bf k})$ would be present on all Fermi surfaces, but be most pronounced on that surface with strongest $k_z$ dispersion, where it has nodes. 
This suggests a close relation between the 3D character of the Fermi surface and gap modulation.

The anisotropy of $\kappa$ shows a strong evolution with Co concentration $x$. At optimal doping, where $T_c$ is maximal,
there are no nodes and $\kappa_0/T$ has the anisotropy of the normal state. 
With increasing $x$, nodes appear and $\kappa_0/T$ acquires a strong anisotropy.
We attribute this to an increase in the gap modulation with $x$, which may explain the strong decrease in the specific heat jump at $T_c$~\cite{Budko2009}
and the change in the power-law temperature dependence of the penetration depth.\cite{Gordon2009a,Gordon2009}
The fact that nodes are located in regions that dominate $c$-axis conduction is consistent with the fact that the penetration depth along the $c$ axis has a linear temperature dependence.\cite{Martin2010}

Horizontal line nodes in Co-Ba122, which would be the result of a strong modulation of the gap along $k_z$ rather than a strong in-plane angular dependence, would reconcile the isotropic azimuthal angular dependence of the gap seen by ARPES with the evidence of nodes or minima from thermal conductivity, NMR relaxation rate and penetration depth measurements in Co-Ba122 and other iron-based superconductors.
A $k_z$ modulation of $\Delta({\bf k})$ should be detectable by ARPES, especially in the overdoped regime where it would be strongest.

Because the nodes go away by tuning $x$ towards optimal doping, we infer that they are `accidental', {\it i.e.} not imposed by symmetry, and so consistent {\it a priori} with any superconducting order parameter, including the $s_{\pm}$ state.\cite{Mazin2008,Kuroki2008,Vorontsov2008}
Although accidental nodes are not a direct signature of the symmetry, the strong modulation of the gap nevertheless reflects an underlying ${\bf k}$ 
dependence of the pairing interaction, and as such the 3D character of the gap function $\Delta({\bf k})$ is an important element in understanding what controls $T_c$ in this family of superconductors.

%%%%%%%%%%%%%%%%%  Acknowledgements

\section{Acknowledgements}

We thank P.~J.~Hirschfeld, V.~G.~Kogan, P.~A. Lee, I.~I.~Mazin, S. Sachdev and T. Senthil for fruitful discussions, 
and J. Corbin for his assistance with the experiments. 
Work at the Ames Laboratory was supported by the US Department of Energy, Office of Basic Energy Sciences under Contract No. DE-AC02-07CH11358.
R.~P. acknowledges support from the Alfred P. Sloan Foundation.
L.~T. acknowledges support from the Canadian Institute for Advanced Research and funding from NSERC, CFI, FQRNT and a Canada Research Chair.

%%%%%%%%%%%%%%%%%%%%%%%%%%%% BIBLIOGRAPHY

%%%% Do not edit the bibliography. Ask to me (JPR) to do it if you want to change something. 
%%%% I'm using the program Jabref to organize the references. 


\begin{references}

\bibitem{Kamihara2008} Y. Kamihara, T. Watanabe, M. Hirano, and H. Hosono,
J.~Amer. Chem. Soc. {\bf 130}, 3296 (2008). 

\bibitem{Zhi-An2008} 
Z.-A. Ren, W. Lu, J. Yang, W. Yi, X.-L. Shen, Z.-C. Li, G.-C. Che, X.-L. Dong, L.-L. Sun, F. Zhou, and Z.-X. Zhao
Chin. Phys. Lett. {\bf 25}, 2215 (2008).

\bibitem{Ishida2009} K. Ishida, Y. Nakai, and H. Hosono, 
J. Phys. Soc. Jpn. {\bf 78}, 062001 (2009).

\bibitem{Mazin2010} I. I. Mazin,  
Nature {\bf 464}, 183 (2010).

\bibitem{Alireza2009} P. L. Alireza, Y. T. C. Ko, J. Gillett, C. M. Petrone, J.~M.~Cole, G. G. Lonzarich, and S. E. Sebastian, 
J. Phys: Condens. Matter {\bf 21}, 012208 (2009).

\bibitem{Rotter2008} M. Rotter, M. Tegel, and D. Johrendt, 
Phys. Rev. Lett. {\bf 101}, 107006 (2008).

\bibitem{Sefat2008} A. S. Sefat, R. Jin, M. A. McGuire, B. C. Sales, D. J. Singh, and D. Mandrus, 
Phys. Rev. Lett. {\bf 101}, 117004 (2008).

\bibitem{Ni2008} N. Ni, M. E. Tillman, J.-Q. Yan, A. Kracher, S. T. Hannahs, S. L. Bud'ko, and P. C. Canfield,
Phys. Rev. B {\bf 78}, 214515 (2008).

\bibitem{Nandi2010} S. Nandi, M. G. Kim, A. Kreyssig, R. M. Fernandes, D.~K.~Pratt, A. Thaler, N. Ni, S. L. Bud'ko, P. C. Canfield, J. Schmalian, R. J. McQueeney, 
and A. I. Goldman,
Phys. Rev. Lett. {\bf 104}, 057006 (2010). 

\bibitem{Fernandes2010} R. M. Fernandes, D. K. Pratt, W. Tian, J. Zarestky, A.~Kreyssig, S. Nandi, M. G. Kim, A. Thaler, N. Ni, P. C. Canfield,
R. J. McQueeney, J. Schmalian, and A. I. Goldman, 
Phys. Rev. B {\bf 81}, 140501 (2010). 

\bibitem{Nakayama2009}  K. Nakayama, T. Sato, P. Richard, Y.-M. Xu, Y. Sekiba, S. Souma, G. F. Chen, J. L. Luo, N. L. Wang, H. Ding, and T. Takahashi,
Europhys. Lett. {\bf 85}, 67002 (2009).

\bibitem{Terashima2009} K. Terashima, Y. Sekiba, J. H. Bowen, K. Nakayama, T.~Kawahara, T. Sato, P. Richard, Y.-M. Xu, L. J. Li, G. H. Cao, Z.-A. Xu, H. Ding, 
and T. Takahashi, 
Proc. Natl. Acad. Sci. U.S.A. {\bf 106}, 7330 (2009).

\bibitem{Samuely2009} P. Samuely, Z. Pribulova, P. Szabo, G. Pristas, S. L. Bud'ko, and P. C. Canfield, 
Physica C {\bf 469}, 507 (2009).

\bibitem{Xu2009} Y.-M. Xu, P. Richard, K. Nakayama, T. Kawahara, Y. Sekiba, T. Qian, M. Neupane, S. Souma, T. Sato, T. Takahashi, H. Luo, H.-H. Wen, G.-F. Chen, N.-L. Wang, Z. Wang, Z. Fang, X. Dai, and H. Ding,
arXiv: 0905.4467. 

\bibitem{Mazin2008} I. I. Mazin, D. J. Singh, M. D. Johannes, and M. H. Du, 
Phys. Rev. Lett. {\bf 101}, 057003 (2008).

\bibitem{Kuroki2008} K. Kuroki, S. Onari, R. Arita, H. Usui, Y. Tanaka, H. Kontani, and H. Aoki, 
Phys. Rev. Lett. {\bf 101}, 087004 (2008).

\bibitem{Vorontsov2008} A. B. Vorontsov, M. G. Vavilov, and A. V. Chubukov,
Phys. Rev. B {\bf 79}, 060508 (2008).

\bibitem{Gordon2009a}  R. T. Gordon, N. Ni, C. Martin, M. A. Tanatar, M. D. Vannette, H. Kim, G. D. Samolyuk, J. Schmalian, S. Nandi, A. Kreyssig, A. I. Goldman, 
J. Q. Yan, S. L. Bud'ko, P. C. Canfield, and R. Prozorov,
Phys. Rev. Lett. {\bf 102}, 127004 (2009).

\bibitem{Gordon2009} R. T. Gordon, C.~Martin, H. Kim, N. Ni, M. A.~Tanatar, J.~Schmalian, I. I.~Mazin, S. L. Bud'ko, P. C. Canfield, and R. Prozorov 
Phys. Rev. B {\bf 79}, 100506 (R) (2009).

\bibitem{Fukazawa2009} H. Fukazawa, Y. Yamada, K. Kondo, T. Saito, Y. Kohori, K. Kuga, Y. Matsumoto, S. Nakatsuji, H. Kito, P. M. Shirage, K. Kihou, N. Takeshita, 
C.-H. Lee, A. Iyo, and H. Eisaki, 
J. Phys. Soc. Jpn. \textbf{78}, 033704 (2009). 

\bibitem{Luo2009}  X. G. Luo, M. A. Tanatar, J.-Ph. Reid, H. Shakeripour, N. Doiron-Leyraud, N. Ni, S. L. Bud'ko, P. C. Canfield, H. Luo, Z. Wang, H.-H. Wen, 
R. Prozorov, and L. Taillefer, 
Phys. Rev. B {\bf 80}, 140503 (R) (2009).

\bibitem{Tanatar2010} M. A. Tanatar, J.-Ph. Reid, H. Shakeripour, X. G. Luo, N. Doiron-Leyraud, N. Ni, S. L. Bud'ko, P. C. Canfield, R. Prozorov, and L. Taillefer, 
Phys. Rev. Lett. {\bf 104}, 067002 (2010).

\bibitem{Dong2010} J. K. Dong, S. Y. Zhou, T. Y. Guan, X. Qiu, C. Zhang, P. Cheng, L. Fang, H. H. Wen, and S. Y. Li, 
Phys. Rev. B {\bf 81}, 094520 (2010).

\bibitem{Martin2010} C. Martin, H. Kim, R. T. Gordon, N. Ni, V. G. Kogan, S. L. Bud'ko, P. C. Canfield, M. A. Tanatar, and R. Prozorov, 
Phys. Rev. B {\bf 81}, 060505 (2010). 

\bibitem{Malaeb2009} W. Malaeb, T. Yoshida, A. Fujimori, M. Kubota, K. Ono, K. Kihou, P. M. Shirage, H. Kito, A. Iyo, H. Eisaki, Y. Nakajima, T. Tamegai, and R. Arita, 
J. Phys. Soc. Jpn. {\bf 78}, 123706 (2009).

\bibitem{Utfeld2010} C. Utfeld, J. Laverock, T. D. Haynes, S. B. Dugdale, J. A. Duffy, M. W. Butchers, J. W. Taylor, S. R. Giblin, J. G. Analytis, J. Chu, I. R. Fisher, 
M. Itou,  and Y. Sakurai, 
Phys. Rev. B {\bf 81}, 064509 (2010). 

\bibitem{Kemper2009} A. F. Kemper, C. Cao, P. J. Hirschfeld, and H.-P. Cheng,  
Phys. Rev. B {\bf 80}, 104511 (2009). 

\bibitem{Analytis2009} J. G. Analytis, R. D. McDonald, J.-H. Chu, S. C. Riggs, A. F. Bangura, C. Kucharczyk, M. Johannes, and I. R. Fisher, 
Phys. Rev. B {\bf 80}, 064507 (2009).

\bibitem{Tanatar2009} M. A. Tanatar, N. Ni, C. Martin, R. T. Gordon, H. Kim, V. G. Kogan, G. D. Samolyuk, S. L. Bud'ko, P. C. Canfield, and R. Prozorov, 
Phys. Rev. B {\bf 79}, 094507 (2009).
 
\bibitem{Tanatar2009a} M. A. Tanatar, N. Ni, G. D. Samolyuk, S. L. Bud'ko, P. C. Canfield, and R. Prozorov,  
Phys. Rev. B {\bf 79}, 134528 (2009). 

\bibitem{Laad2009} M. S. Laad and L. Craco, 
Phys. Rev. Lett. {\bf 103}, 017002 (2009). 

\bibitem{Graser2010} S. Graser, A. F. Kemper, T. A. Maier, H.-P. Cheng, P. J. Hirschfeld, and D. J. Scalapino, 
arXiv:1003.0133. 

\bibitem{Chi2009} S. Chi, A. Schneidewind, J. Zhao, L. W. Harriger, L. Li, Y. Luo, G. Cao, Z. Xu, M. Loewenhaupt, J. Hu, and P. Dai,
Phys. Rev. Lett. {\bf 102}, 107006 (2009). 

\bibitem{Hirschfeld1988} P. J. Hirschfeld, P. Wolfle, and D. Einzel,
Phys. Rev. B {\bf 37}, 83 (1988).

\bibitem{Graf1996} M. J. Graf, S.-K. Yip, J. A. Sauls, and D. Rainer,
Phys. Rev. B {\bf 53}, 15147 (1996).

\bibitem{Durst2000} A. C. Durst and P.A. Lee, 
Phys. Rev. B {\bf 62}, 1270 (2000).

\bibitem{Taillefer1997} L. Taillefer, B. Lussier, R. Gagnon, K. Behnia, and H. Aubin,
Phys. Rev. Lett. {\bf 79}, 483 (1997). 

\bibitem{Suzuki2002} M. Suzuki, M. A. Tanatar, N. Kikugawa, Z. Q. Mao, Y. Maeno, and T. Ishiguro,
Phys. Rev. Lett. {\bf 88}, 227004 (2002). 

\bibitem{Shakeripour2009a} H. Shakeripour, C. Petrovic, and L. Taillefer, 
New J. Phys. {\bf 11}, 055065 (2009).

\bibitem{Shakeripour2007} H. Shakeripour, M. A. Tanatar, S. Y. Li, C. Petrovic, and L. Taillefer, 
Phys. Rev. Lett. {\bf 99}, 187004 (2007). 

\bibitem{Kano2009} M. Kano, Y. Kohama, D. Graf, F. Balakirev, A. S. Sefat, M. A. Mcguire, B. C. Sales, D. Mandrus, and S. W. Tozer,
J. Phys. Soc. Jpn.  {\bf 78}, 084719 (2009).

\bibitem{Sutherland2003} 
M. Sutherland, D. G. Hawthorn, R. W. Hill, F. Ronning, S. Wakimoto, H. Zhang, C. Proust, E. Boaknin, C. Lupien, L. Taillefer, 
R. Liang, D. A. Bonn, W. N. Hardy, R. Gagnon, N. E. Hussey, T. Kimura, M. Nohara, and H. Takagi,
Phys. Rev. B {\bf 67}, 174520 (2003).

\bibitem{Tanatar2010a} M. A. Tanatar, N. Ni, S. L. Bud'ko, P. C.~Canfield, and R. Prozorov, 
Supercond. Sci. Technol. (to be published).

\bibitem{Hawthorn2007} D. G. Hawthorn, S. Y. Li, M. Sutherland, E. Boaknin, R. W. Hill, C. Proust, F. Ronning, M. A. Tanatar, J. Paglione, and L. Taillefer,
Phys. Rev. B {\bf 75}, 104518 (2007).

\bibitem{Li2008} S. Y. Li, J.-B. Bonnemaison, A. Payeur, P. Fournier, C. H. Wang, X. H. Chen, and L. Taillefer,  
Phys. Rev. B {\bf 77}, 134501 (2008).

\bibitem{Ding2009} L. Ding, J. K. Dong, S. Y. Zhou, T. Y. Guan, X. Qiu, C. Zhang, L. J. Li, X. Lin, G. H. Cao, Z. A. Xu, and S. Y. Li, 
New J. Phys. {\bf 11}, 093018 (2009).

\bibitem{Mishra2009} V. Mishra, A. Vorontsov, P. J. Hirschfeld, and I. Vekhter, 
Phys. Rev. B {\bf 80}, 224525 (2009).

\bibitem{Kogan2009} V. G. Kogan, 
Phys. Rev. B {\bf 80}, 214532 (2009),

\bibitem{Shakeripour2009} H. Shakeripour, M. A. Tanatar, C. Petrovic, and L. Taillefer, 
arXiv:0902.1190.

\bibitem{Proust2002} C. Proust, E. Boaknin, R.W. Hill, L. Taillefer, and A. P. Mackenzie, 
Phys. Rev. Lett. {\bf 89}, 147003 (2002).


\bibitem{Budko2009} S. L. Bud'ko, N. Ni, and P. C.~Canfield, 
Phys. Rev. B {\bf 79}, 220516 (2009).

\bibitem{Nakamae2001} S. Nakamae, K. Behnia, L. Balicas, F. Rullier-Albenque, H. Berger, and T. Tamegai,
Phys. Rev. B {\bf 63}, 184509 (2001).

\bibitem{Kubert1998} C. Kubert and P. J. Hirschfeld, 
Phys. Rev. Lett. {\bf 80}, 4963 (1998).

\bibitem{Boaknin2003} E. Boaknin, M. A. Tanatar, J. Paglione, D. G. Hawthorn, F. Ronning, R. W. Hill, M. Sutherland, L. Taillefer, J. Sonier, S. M. Hayden, 
and J. W. Brill, 
Phys. Rev. Lett. {\bf 90}, 117003 (2003).

\bibitem{Dong2010a} 
J. K. Dong, S. Y. Zhou, T. Y. Guan, H. Zhang, Y. F. Dai, X. Qiu, X. F. Wang, Y. He, X. H. Chen, and S. Y. Li,
Phys. Rev. Lett. {\bf 104}, 087005 (2010).

\bibitem{Bouquet2002} F. Bouquet, Y. Wang, I. Sheikin, T. Plackowski, A. Junod, S. Lee, and S. Tajima, 
Phys. Rev. Lett. {\bf 89}, 257001 (2002).

\bibitem{Gubser1973} D. U. Gubser and J. E. Cox, 
Phys. Rev. B {\bf 7}, 4118 (1973).

\bibitem{Mazin2009} I. I. Mazin and J. Schmalian, 
Physica C {\bf 469}, 614 (2009). 

\bibitem{Zhitomirsky2001a} M. E. Zhitomirsky and T. M. Rice, 
Phys. Rev. Lett. {\bf 87}, 057001 (2001).


\end{references}
\end{document}